\documentclass[twocolumn,amsmath,amssymb,pre]{revtex4}

\usepackage{mathtools}
\usepackage{graphicx}
\usepackage{dcolumn}
\usepackage{bm}
\usepackage{makecell}
\usepackage{xcolor}
\usepackage{enumerate}
\usepackage[caption=false]{subfig}
\usepackage{float}
\usepackage[unicode=true,pdfusetitle,
 bookmarks=true,bookmarksnumbered=false,bookmarksopen=false,
 breaklinks=false,pdfborder={0 0 1},backref=false,colorlinks=true]
 {hyperref}
\hypersetup{linkcolor=blue,urlcolor=blue,citecolor=blue}

\newcommand{\be}{\begin{equation}}
\newcommand{\ee}{\end{equation}}
\newcommand{\bd}{\begin{displaymath}}
\newcommand{\ed}{\end{displaymath}}
\newcommand{\BE}{\begin{eqnarray}}
\newcommand{\EE}{\end{eqnarray}}

\newcommand{\bx}{\ensuremath{\mathbf{x}}}

\newcommand{\bn}{\ensuremath{\mathbf{n}}}

\newcommand{\avg}[1]{\left\langle{#1}\right\rangle}

\begin{document}

\title{Local and global ordering dynamics in multi-state voter models}

 \author{Luc\'ia Ramirez}
\email{luciaramirez@ifisc.uib-csic.es}
\affiliation{Instituto de F\'isica Interdisciplinar y Sistemas Complejos, IFISC (CSIC-UIB), Campus Universitat Illes Balears, E-07122 Palma de Mallorca, Spain}

\affiliation{Departamento de F\'{\i}sica, Instituto de F\'{\i}sica Aplicada, Universidad Nacional de San Luis-CONICET, Ej\'ercito de Los Andes 950, D5700HHW, San  Luis, Argentina}

\author{Maxi San Miguel}
\email{maxi@ifisc.uib-csic.es}
 \affiliation{Instituto de F\'isica Interdisciplinar y Sistemas Complejos, IFISC (CSIC-UIB), Campus Universitat Illes Balears, E-07122 Palma de Mallorca, Spain}

 \author{Tobias Galla}
\email{tobias.galla@ifisc.uib-csic.es}
 
\affiliation{Instituto de F\'isica Interdisciplinar y Sistemas Complejos, IFISC (CSIC-UIB), Campus Universitat Illes Balears, E-07122 Palma de Mallorca, Spain}

\begin{abstract}
We investigate the time evolution of the density of active links and of the entropy of the distribution of agents among opinions in multi-state voter models with all-to-all interaction and on uncorrelated networks. Individual realisations undergo a sequence of eliminations of opinions until consensus is reached.  After each elimination the population remains in a meta-stable state. The density of active links and the entropy in these states varies from realisation to realisation. Making some simple assumptions we are able to analytically calculate the average density of active links and the average entropy in each of these states. We also show that, averaged over realisations, the density of active links decays exponentially, with a time scale set by the size and geometry of the graph, but independent of the initial number of opinion states. The decay of the average entropy is  exponential only at long times when there are at most two opinions left in the population. Finally, we show how meta-stable states comprised of only a subset of opinions can be artificially engineered by introducing precisely one zealot in each of the prevailing opinions.

\end{abstract}
\maketitle

\section{Introduction}
One of the most popular classes of models of opinion dynamics is that of so-called voter model (VM)~\cite{clifford,liggett1,holey,castellano_fortunato,redner}. VMs describe populations of individuals, who are each characterised by their discrete opinion state, and where the principal mechanism of change is imitation (i.e., one individual copies the opinion state of another individual). VMs are not only a paradigmatic model of opinion formation, they are also of interest in statistical physics. They operate out of equilibrium, and have absorbing states and certain symmetries, defining an interesting universality class~\cite{liggett1,dornic}.

In the most simple version of the VM each individual can take one of two opinion states. Individuals are assumed to reside on the nodes of an interaction network (this includes the special case of all-to-all interaction). At each step an individual is chosen at random and then copies the state of one randomly chosen neighbour on the network. This simple system has $\mathbb{Z}_2$-symmetry, and two absorbing `consensus' states (if all individuals hold the same opinion, no further change is possible).  The features of most interest to physicists include the time required to reach absorption, and the coarsening dynamics during the process leading to consensus \cite{castellano_fortunato, dornic, redner,sood_redner,sood,ben,suchecki,vazquez_eguiluz}.

 Many variants of the voter model have been introduced in order to capture different features of social interaction. Examples are the so-called noisy voter model in which individuals can change opinion spontaneously \cite{carro, peralta_carro, herrerias, vazquez}, or populations including so-called `zealots'. These are individuals who are less prone to changing opinion than regular agents, or who never change opinion  \cite{nagi,mobilia3,mobilia4, khalilz}. A further variation, which we focus on here, is the so-called multi-state voter model (MSVM) \cite{starnini}. These are voter models in which there are more than two possible opinion states, and consequently multiple absorbing states. The path to one of these states involves a sequence of successive extinctions of opinions. One main distinction is between models in which the different opinion states are ordered in some way (representing e.g. the political spectrum) versus models in which all states are equivalent. We here focus on the latter case.

Existing literature on MSVM with equivalent states includes in particular work on consensus and extinction times ~\cite{baxter, starnini, pickering}. For the case of all-to-all interaction the authors of Ref.~\cite{starnini} derived analytical expressions for objects such as the mean consensus time, and the mean number of different states in the population as a function of time.  Ref.~\cite{starnini} also contains numerical studies of the model in low-dimensional lattices.

The authors of Ref.~\cite{pickering}, among other results, further provided closed-form expressions for all moments of the consensus time for uniform initial distributions on the all-to-all interaction. Baxter {\em et al.} \cite{baxter} obtain a solution for a model describing neutral genetic drift at a single locus with multiple alleles. This model, while set up in a biological context, is mathematically very similar to the MSVM in an all-to-all geometry, and the bulk of the ideas and results therefore carry over. It should be noted that the authors of \cite{baxter} work in the diffusion approximation.

Many of the results in the existing literature concern quantities such as the consensus time, the time until the extinction of the different opinions, or the properties of the stationary state (the latter becomes non-trivial if spontaneous opinion changes are added to the dynamics, as this removes the absorbing states). 

The question how consensus is reached in MSVMs, and what the coarsening process before absorption looks like, on the other hand, appear to have received relatively little attention. In this work, we therefore study the VM with a general number of initial opinion states, and with a focus on the dynamics before consensus is reached. 
 
We focus on the cases of a complete graph (all-to-all interaction), uncorrelated networks such as Erd\"os--Renyi  graphs, and scale-free networks. Throughout our paper scale-free networks are generated using the Barab\'asi--Albert growth process of preferential attachment \cite{barabasi}, and have a degree distribution which decays as $p(k)\sim k^{-3}$. We will simply refer to these as Barab\'asi--Albert networks. The assumption of an uncorrelated network is then valid \cite{suchecki,vazquez_eguiluz}.

We focus on two key quantities in our analysis. One is the familiar `density of active links' \cite{dornic,cast3,cast,suchecki,vazquez_eguiluz,carro,pugliese}, that is the proportion of connected pairs of agents that do not share the same opinion. This quantity characterises the organisation of individuals on the graph. A low density of active links indicates the presence of domains of individuals of the same state (neighbours tend to be in the same opinion states). The pattern is more scattered if the density of active links is high \cite{kauhanen}. Additionally, we look at the entropy of the distribution of agents across the different opinion states. This is a global measure of order (it does not make use of pairs of neighbours), indicating how dispersed the individuals are across the different opinions. 

The time evolution of the density of active links in the two-state model has been studied on complete graphs \cite{cast,ben}, and on uncorrelated graphs~\cite{suchecki,pugliese,vazquez_eguiluz, peralta_carro,carro}. Averaged over an ensemble of realisations, an exponential decay of the mean density of active links with time is here typically found. The decay time  is proportional to the population size, $N$, for complete graphs and Erd\"os-Renyi networks  \cite{sood,cast}. On  Barab\'asi--Albert networks the time scale is proportional to $N/\ln N$ \cite{sood_redner}.

The behaviour in individual realisations is quite different. Both in the all-to-all scenario and on graphs one finds that single runs of the two-state VM typically settle to quasi-stable `plateau' of the density of active links, before a sudden fluctuation then takes the system to consensus \cite{suchecki,klemm}. The density of active links at the plateau differs in the cases of all-to-connectivity or networks. It also depends on the initial proportion of agents in the two opinion states \cite{vazquez_eguiluz}.   
 
In the two-state model consensus is reached after a single extinction of an opinion. The multi-state model on the other hand undergoes a sequence of extinctions. Our main objective is to study how this affects the time evolution of the density of active links, and of the entropy. We address this both at the ensemble level (i.e., as an average over realisations), and on the level of individual runs. 

The remainder of the paper is set out as follows. In Sec.~\ref{sec:model} we define the model. Sec.~\ref{sec:ensemble} then focuses on the time evolution of the density of active links and of the entropy at the level of an ensemble average. The phenomenology of individual realisations is studied in more detail in Sec.~\ref{sec:individual_realisations}. We find a sequence of meta-stable states, and characterise some of the statistical features of these states. We then use this to  establish how the ensemble-level behaviour can be understood from that of single realisations. In Sec.~\ref{sec:zealots} we then proceed to show that zealots can be used to `engineer' steady states of mixed opinions similar to the meta-stable states found in individual realisations in Sec.~\ref{sec:individual_realisations}. Finally,
Sec.~\ref{sec:individual_realisations} contains a summary and our conclusions.
\section{Model definitions}\label{sec:model}

\subsection{Setup and interaction network}
The model describes $N$ individuals, who can each be in one of $M$ discrete states or opinions. We label individuals $i=1,\dots,N$ and states $\alpha=1,\dots,M$. During the course of the dynamics each individual can interact with its nearest neighbours on a static network. We use the notation $c_{ij}$ for the adjacency matrix of the undirected interaction network. We have $c_{ij}=c_{ji}=1$ if individuals $i$ and $j$ are neighbours, and $c_{ij}=c_{ji}=0$ otherwise. We also set $c_{ii}=0$. We will use the notation $j\in i$ to indicate that $j$ is among the neighbours of $i$. We write $k_i$ for the degree of the node representing individual $i$, i.e., $k_i=\sum_j c_{ij}$. The total number of links in the graph is $E=\sum_{i<j} c_{ij}$. In the complete graph ($c_{ij}=1$ for all $i\neq j$) one has $E=N(N-1)/2$.

\subsection{Dynamics and transition rates}
The variable $s_i(t)\in\{1,\dots,M\}$ represents the state of individual $i$ at time $t$. At the start of the dynamics ($t=0$) the states of all individuals $s_i(t=0)$ are initialised. Different initial conditions can here be chosen. We typically consider \textit{homogeneous} initial conditions \cite{starnini}, that is a configuration in which the same number of agents in each opinion state are randomly distributed in the nodes of the network. For the model on networks, these individuals are placed on the graph at random. 

The dynamics then proceeds through pairwise interactions between individuals. An interaction of individual $i$ with individual $j\in i$ consists of an imitation process, i.e., $i$ copies the opinion state of $j$. It is important to notice that, although the interaction network is undirected, each interaction event is directed. An interaction of $i$ with $j$ is not the same as an interaction of $j$ with $i$. 

The rates with which interactions between agent $i$ and $j$ occur are given by
\be 
\label{eq:rate}
T_{ij}=\frac{c_{ij}}{k_i}=\left\{\begin{array}{cl}1/k_i~~ & j\in i \\ ~ & ~ \\ 0 & j\notin i\end{array}.\right.
\ee
 
In this setup time is continuous, and measured in units of Monte Carlo steps (`generations' in the language of population dynamics), i.e. ${\cal O}(N)$ imitation events take place in the population per unit time.
 
We denote the number of individuals holding opinion $\alpha$ by $n_\alpha$, and we write $\bn(t)=[n_1(t),\dots, n_M(t)]$. We also introduce $x_\alpha = n_\alpha/N$ as the fraction of individuals in opinion state $\alpha$.

\subsection{Density of active links}

At each point in time the links on the interaction network can be grouped into links that are `active' or `inactive' respectively. A link $(i,j)$ is said to be active when the two nodes at its ends are in different states ($s_i\neq s_j$), otherwise the link is inactive.

It is further useful to introduce the fraction of links of type $\alpha\beta$, where $\alpha,\beta\in\{1,\dots,M\}$. This is the fraction of links which have an agent in state $\alpha$ at one end, and an agent in state $\beta$ at the other end. Suppressing the time-dependence of the $\{s_i(t)\}$ we then have
\be\label{eq:rhoalphabeta}
\rho_{\alpha\beta}=\frac{1}{E}\sum_{i<j} c_{ij} \left(\delta_{s_i,\alpha}\delta_{s_j,\beta}+\delta_{s_j,\alpha}\delta_{s_i,\beta}\right)
\ee
for $\alpha\neq\beta$, and with $\delta$ the Kronecker delta. We recall that $E$ is the total number of links in the graph. The total density of active links in the system is then
\be\label{eq:rho}
\rho=\sum_{\alpha<\beta} \rho_{\alpha\beta}=\frac{1}{E}\sum_{i<j} c_{ij} (1-\delta_{s_i,s_j}).
\ee

The overall rate of events in the population leading to state changes is proportional to the fraction of links that are active (an imitation process involving individuals connected by an inactive link will not result in any opinion change). The density of active links therefore characterises the amount of (potential) `activity', and indicates how far the system is from reaching an absorbing state. 

The density of active links can also be seen as a measure of disorder in the configuration of opinions on the interaction network. Consider a particular configuration $\bn$ of individuals in the different opinion states. If the individuals were located at random nodes on the graph with no particular order, then the probability that a randomly chosen link is active is $\rho_{\rm random}=2\sum_{\alpha<\beta}n_\alpha n_\beta/[N(N-1)]$. This is also the fraction of active links on a complete graph, $\rho_{\rm CG}(\bn)$, given the $\{n_\alpha\}$. If there is order in the network (i.e., the neighbours of a node in state $\alpha$ also tend to be in state $\alpha$), then the density of links will be lower than $\rho_{\rm CG}(\bn)$. At an absorbing state one has complete order, $\rho = 0$.  Thus, $\rho(\bn)$ indicates the amount of disorder in a configuration of the networked system. The behaviour of this quantity in time can be used to characterise the coarsening dynamics.  We stress that the density of active links arises from a local definition of disorder (the state of an individual is compared with that of its neighbours).

\subsection{Entropy}
We now introduce a second measure of disorder, namely the entropy of a particular configuration. This is defined as
\be
S=-\sum_\alpha x_\alpha \ln\,x_\alpha,
\ee
where we recall that $x_\alpha=n_\alpha/N$. This definition makes no use of neighbourhood relationships between nodes. Hence entropy as a  measure of disorder has a global character.

States of maximum entropy are those for which $x_\alpha=1/M$ for all $\alpha$, i.e., states with equally many individuals in each opinion state. This leads to $S=\ln\, M$. If the system has reached consensus ($x_\alpha=1$ for one value of $\alpha$, and $x_\beta=0$ for all $
\beta\neq \alpha$), we have $S=0$. This is the state of maximal order.

\subsection{Master equation for the model on a complete graph}\label{sec:simplify_CG}
In an all-to-all geometry all individuals are equivalent in terms of their position on the graph, and the system is therefore fully specified by $\bn=(n_1,\dots,n_M)$. The rates in Eq.~(\ref{eq:rate}) become $T_{ij}=1/(N-1)$ for all $i\neq j$. This in turn means that the total rate with which individuals in state $\alpha$ in the population are converted to individuals in state $\beta$ is 
\be\label{eq:rates_aa}
{\cal T}_{\alpha\to\beta}(\bn)=\frac{n_\alpha n_\beta}{N-1}.
\ee

The dynamics of the system is then described by the master equation
\be\label{eq:master_all_to_all}
\frac{d}{dt}P(\bn)=\sum_{\alpha\neq\beta} (E_\alpha E_\beta^{-1}-1)[{\cal T}_{\alpha\to\beta}(\bn)P(\bn)],
\ee
where $E_\alpha$ is the `raising operator' acting on functions of $\bn$ by increasing the argument $n_\alpha$ by one, $E_\alpha f(\bn)=f(n_1,\dots, n_\alpha+1,\dots, n_M)$. 

\section{Characterisation of the coarsening process at the level of the ensemble average}\label{sec:ensemble} 
 
To set the scene we will first focus on the time evolution of averaged quantities. By this we mean an  average over realisations of the stochastic voter model dynamics, each realisation with a different initial condition. We will refer to this as an `ensemble average', and use angle brackets $\avg{\cdots}$ to describe it.

\subsection{Evolution of the average density of links}

\subsubsection{Complete Graph}

In the basic case of all-to-all interactions one has, for any function $f(\bn)$,
\be
\frac{d}{dt}\avg{f}=\sum_\bn f(\bn)\frac{d}{dt}P(\bn),
\ee
with $\frac{d}{dt}P(\bn)$ as in Eq.~(\ref{eq:master_all_to_all}).

From Eqs.~(\ref{eq:rate}) it is also clear that the rate for events converting individuals from state $\alpha$ to state $\beta$ is the same as that for the reverse event. Therefore, the ensemble average of $x_\alpha$ is constant in time,
\be \label{eq:dt}
\frac{d}{dt} \avg{x_\alpha} =0.
\ee

If an individual of type $\alpha$ adopts opinion $\beta$ in an event, then we have $n_\alpha\to n_\alpha-1$ and $n_\beta\to n_\beta+1$. Given that $\rho_{\alpha\beta}=2n_\alpha n_\beta/[N(N-1)]$ this means that $\rho_{\alpha\beta}$ changes by $2(n_\alpha-n_\beta-1)/[N(N-1)]$. We therefore have
\BE
\frac{d}{dt}\avg{\rho_{\alpha\beta}}&=&\frac{2}{N(N-1)}\bigg<{\cal T}_{\alpha\to \beta}(\bn)\times (n_\alpha-n_\beta-1)\nonumber \\
&&+{\cal T}_{\beta\to\alpha}(\bn)\times(n_\beta-n_\alpha-1)\bigg>\nonumber 
\\
&=&-\frac{2}{N-1}\avg{\rho_{\alpha\beta}}.
\EE
We conclude
\be\label{eq:exp}
\avg{\rho_{\alpha\beta}(t)}=\avg{\rho_{\alpha\beta}(0)}e^{-t/\tau},
\ee
with a time scale $\tau$ given by~\cite{cast}
\be\label{eq:tau_all_to_all}
\tau=\frac{N-1}{2}.
\ee

From this and using Eq.~(\ref{eq:rho}) we have
\be\label{eq:exp2}
\avg{\rho}=\avg{\rho(0)} e^{-t/\tau}.
\ee

We note that $\tau$ is independent of the number of opinion states $M$.
\begin{figure}
\centering
\includegraphics[width=0.95\columnwidth]{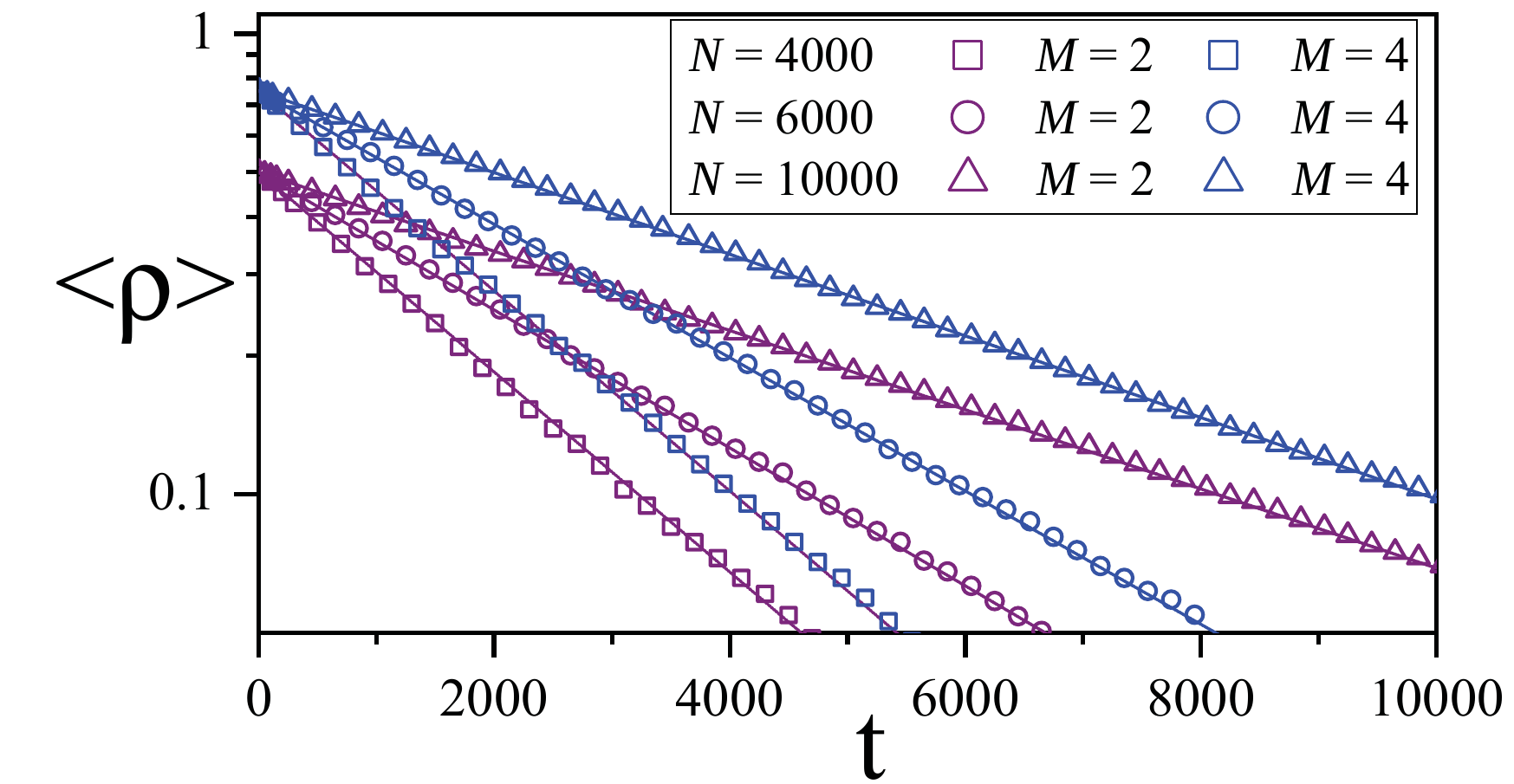}
\caption{Average density of links in the MSVM with all-to-all interaction and homogeneous initial conditions (see text) for different values of initial number of opinions $M$ and different system sizes $N$. Symbols show results from numerical simulations, averaged over $5000$ realisations. Lines are the analytical prediction in Eq.~(\ref{eq:exp2}). \label{fig:expCG}}
\end{figure}

For homogeneous initial conditions ($n_\alpha=N/M$ for all $\alpha$), we have 
\be\label{eq:rho_hom}
\avg{\rho(0)}=\frac{(M-1)N}{(N-1)M}\approx {1-\frac{1}{M}},
\ee
where the approximation applies for $N\gg 1$.

\begin{figure}
\centering
\includegraphics[width=0.95\columnwidth]{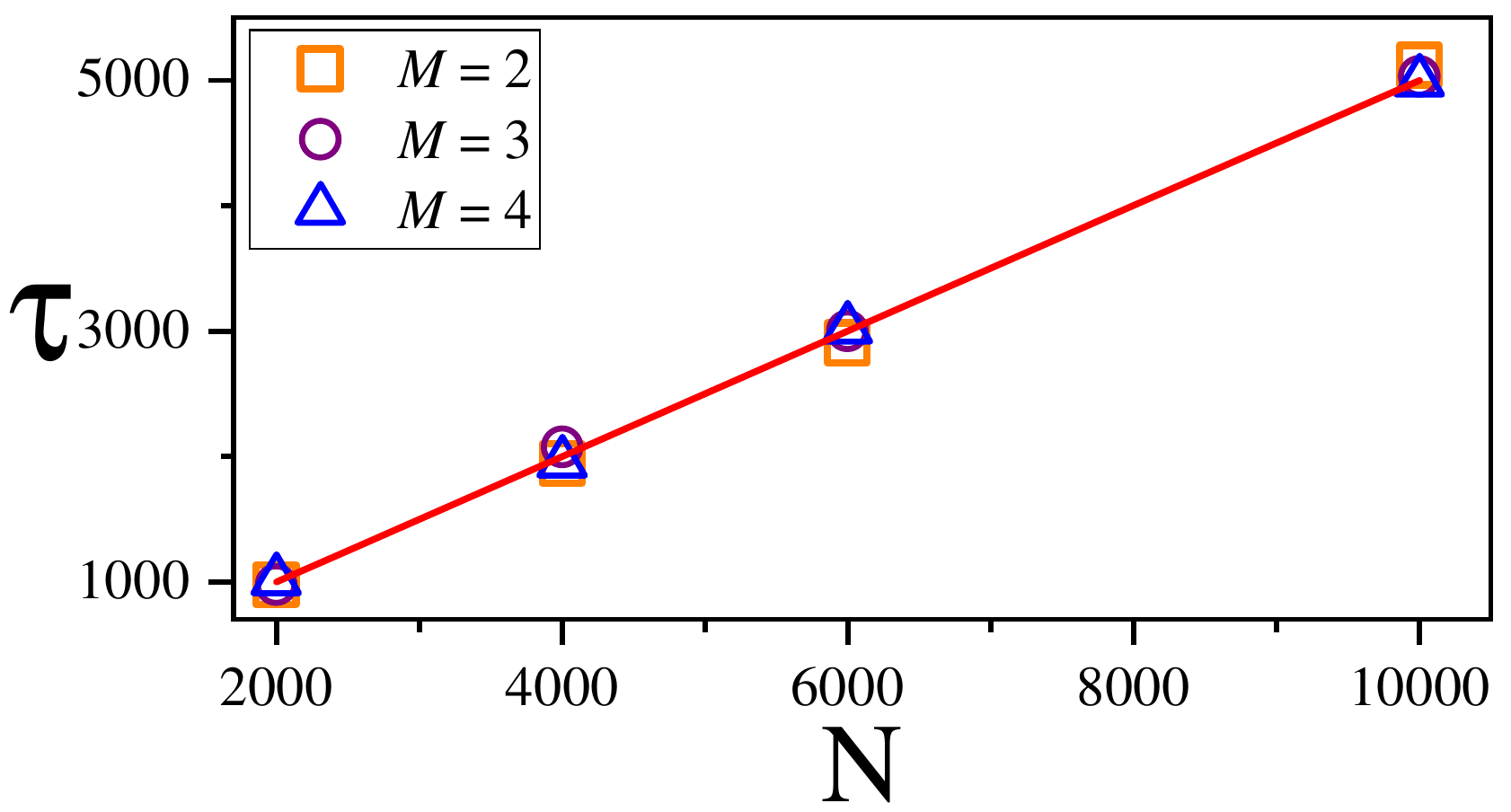}
\caption{Decay time scale $\tau$ in the MSVM on a complete graph for different values of initial number of opinions $M$, as a function of the system size $N$. Markers are obtained from fitting an exponential curve to simulation data for $\avg{\rho(t)}$. The solid line shows  Eq.~(\ref{eq:tau_all_to_all}).  \label{fig:tau_cg}}
\end{figure}

In Fig.~\ref{fig:expCG} we confirm the validity of Eq.~(\ref{eq:exp2}) in simulations. The expression in Eq.~(\ref{eq:tau_all_to_all}) is verified in Fig.~\ref{fig:tau_cg}.

\subsubsection{Pair approximation for the two-state model on uncorrelated networks}

In a networked geometry it is not straightforward to obtain closed laws for the time evolution of macroscopic average quantities such as the density of active links. This is because the state of the system is no longer fully described by the numbers $n_1,\dots,n_M$. The correlations that build up between nodes in the network are one key element distinguishing the voter process on networks from that with all-to-all interaction.

Analytical progress for the model on networks is possible as an approximation. One prominent approach is the so-called `pair approximation' \cite{dickman,pugliese,vazquez_eguiluz, peralta_carro}, capturing correlations between nearest neighbour nodes, but not between nodes which are further apart on the graph. The pair approximation is known to capture the behaviour on uncorrelated networks to a good accuracy \cite{pugliese,vazquez_eguiluz, peralta_carro,carro}. These networks can have an arbitrary degree distribution, but the degrees of nodes are uncorrelated, including the degrees of the nearest neighbour~\cite{dor} (the probability that any two nodes are connected therefore only depends on their degrees). This includes Erd\"os--Renyi and Barab\'asi--Albert networks.   

Within the pair approximation the following expression can be found for the average density of active links in the VM with two opinion states \cite{vazquez_eguiluz, peralta_carro},
\be\label{eq:rho_network}
\avg{\rho(t)}=2\avg{x_1(0)[1-x_1(0)]}\frac{\overline k-2}{\overline k-1}e^{-t/\tau}~\mbox{for}~ t>t^\star,
\ee
where $\overline k$ is the mean degree of nodes in the network. The quantity $x_1(0)$ is the initial fraction of individuals in opinion state $1$. If this fraction is fixed then the average $\avg{\cdots}$ on the right-hand side of Eq.~(\ref{eq:rho_network}) can be removed. Within the pair approximation the time scale $\tau$ is given by \cite{vazquez_eguiluz}
\be\label{eq:tau_network}
\tau=\frac{(\overline k-1)\overline{k}^2 N}{2(\overline{k}-2)\overline{k^2}}.
\ee

 In this expression, $\overline{k^2}$ is the second moment of the degree distribution of the interaction network.
 
It is important to note that Eq.~(\ref{eq:rho_network}) is only valid after a short transient of duration $t^*$. During this transient, the average density of active links reduces from its initial value $2\avg{x_1(0)[1-x_1(0)]}$ by a factor of $(\overline k-2)/(\overline k-1)$. Unlike the decay time $\tau$, the time scale $t^\star$ does not increase with $N$, i.e.,  we have $t^\star={\cal O}(N^0)$ \cite{vazquez_eguiluz}.

The time evolution of the average density of links in the two-state VM in Eq.~(\ref{eq:rho_network}) consists of three main factors: (i) the initial average density of active links, $2\avg{x_1(0)[1-x_2(0)]}$, one would obtain in an all-to-all geometry for the same distribution of initial proportions of agents in the two opinion states, (ii) a factor $(\overline k -2)/(\overline k -1)$ accounting for the network geometry, and (iii) exponential decay.

\subsubsection{Pair approximation for the MSVM}\label{sec:pa_subsec}
We will now use a simple argument to show that this general structure carries over to the multi-state model. The only change required is to adapt the expression for the initial density of active links to the case of multiple opinion states, given initial proportions of agents in these different states.

The argument is based on the fact that the dynamics of one single opinion in the multi-state voter model can be understood from a reduction to a two-state model. This was first proposed in the context of the VM in Ref.~\cite{redner} and later used also in Ref.~\cite{herrerias}. The same ideas can also be found in earlier work by Kimura and Littler in the field of genetics \cite{kimura,littler}. 

If the focus is on the dynamics of one single opinion (say $\alpha=1$) then it is not necessary to resolve the different other opinion states. Instead what is relevant for an individual to change out of state $\alpha=1$ is that they interact with an individual in {\em any} of the other states. Similarly, an individual changes into state $\alpha=1$ if they were previously in any other state and interact with an individual in state $1$. For the purposes of studying individuals in state $\alpha=1$ it is not required to know what these other states were. This means that all other opinions ($\beta=2,\dots,M$) can be amalgamated into one single opinion state. This then reduces the model to a two-state voter process, one state represents opinion $\alpha=1$ and the second state stands for `all other opinions'. We label these states as $+$ and $-$ respectively. The dynamical rules in Eq.~(\ref{eq:rates_aa}) are such that this reduced models follows the dynamics of a two-state voter model.

Suppose now, we start the multi-state voter process from homogeneous initial conditions, $x_\alpha(0) = 1/M$ for $\alpha=1,\dots,M$, and focus on a particular opinion. We then have $x_+(0)=1/M$, and $x_-(0)=1-1/M$. The reduced model is therefore a two-state voter model with inhomogeneous initial conditions. 

Using Eq.~(\ref{eq:rho_network}) we have the following average density of active links in this reduced model,
\be\label{eq:rho_reduced}
\avg{\rho_{\rm red}(t)} = 2\frac{1}{M}\left(1-\frac{1}{M}\right)\frac{\overline k-2}{\overline k-1}e^{-t/\tau}.
\ee

We note that this is the average fraction of links  between states $+$ and $-$ in the reduced model. In the original model (before the reduction) this corresponds to the fraction of links connecting an individual in state $\alpha=1$ with an individual in {\em any} other state $\beta\neq 1$, i.e., $\rho_{\rm red}=\sum_{\beta>1} \rho_{1,\beta}$. Carrying out the ensemble average, and exploiting the symmetry between states in the MSVM with homogeneous initial conditions we have $\avg{\sum_{\beta>1} \rho_{1,\beta}}=\avg{\sum_{\beta\neq\alpha}
\rho_{\alpha,\beta}}$ for any fixed $\alpha$. Hence, the resulting average density of active links in the MSVM is
\be
\avg{\rho(t)}=\frac{1}{2}\sum_{\alpha\neq \beta} \avg{\rho_{\alpha\beta}(t)}=\frac{M}{2}\avg{\rho_{\rm red}(t)}.
\ee
(In this expression both $\alpha$ and $\beta$ are summed over). Using Eq.~(\ref{eq:rho_reduced}) we then obtain
\be\label{eq:rho_a_network}
\avg{\rho(t)}=\xi(M,\overline k) e^{-t/\tau}~\mbox{for}~t>t^\star,
\ee
with
\be\label{eq:xi}
\xi(M,\overline k)\equiv \left(1-\frac{1}{M}\right)\frac{\overline k-2}{\overline k-1},
\ee
for the multi-state voter model with homogeneous initial conditions. The decay time $\tau$ is as in Eq.~(\ref{eq:tau_network}).  As before, the exponential law in Eq.~(\ref{eq:rho_a_network}) is only valid after a short initial transient of duration $t^\star={\cal O}(N^0)$.

Similar to the two-state model, we notice that the pre-factor $\xi(M,\overline k)$ in Eq.~(\ref{eq:rho_a_network}) is made up of the network-specific factor $(\overline k-2)/(\overline k-1)$, and the density of active links $1-1/M$ in Eq.~(\ref{eq:rho_hom}), resulting from a configuration in which equally many individuals hold each opinion ($x_\alpha=1/M~\forall\alpha$), and in which these individuals are placed on the network at random. The network-specific factor only depends on the mean degree.

\subsubsection{Test against simulations and metastable state in the coarsening dynamics}

We now test the predictions of the previous section in numerical simulations. These were carried out on uncorrelated networks with different degree distributions, specifically the Erd\"os--Renyi and Barab\'asi--Albert ensembles.

\begin{figure} \includegraphics[width=1\columnwidth]{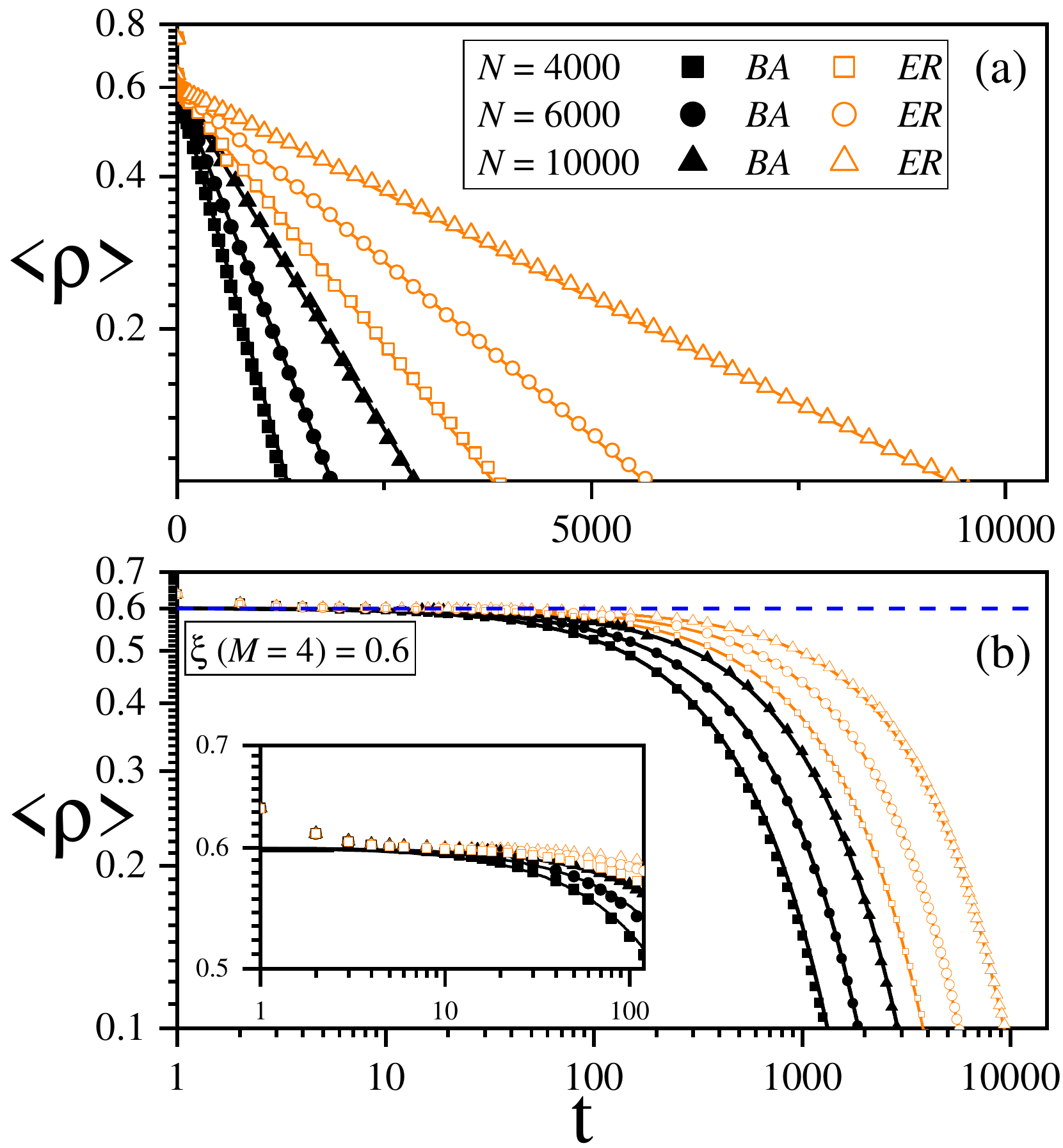}
\caption{(a) Time-evolution of the average density of active links in the multi-state voter model with $M=4$ for different systems sizes $N$. Markers show results from simulations, started from random homogeneous initial conditions, and averaged over $5000$ realisations. We show data for Erd\"os--Renyi graphs (ER, orange open symbols) and for Barab\'asi--Albert networks (BA, black full symbols). All networks have mean degree $\overline k = 6$.  Solid lines are the analytical prediction in Eq.~(\ref{eq:rho_a_network}). Panel (a) is on linear-log scale. In panel (b)  we show the same data on double logarithmic scale, the density of active links shows a plateau  $\xi\approx 0.6$, in-line with Eq.~(\ref{eq:xi}). The lifetime of the plateau is finite for finite populations and increases with $N$.   The inset provides a further zoom-in, highlighting the initial decay of the density of active links from its initial value to $\xi=0.6$. 
\label{fig:rho_net}} \end{figure}

We first verify the validity of Eq.~(\ref{eq:rho_a_network}). We show the time-evolution of the average density of active links, $\avg{\rho(t)}$, from simulations in Fig.~\ref{fig:rho_net}, along with the analytical predictions from the pair approximation. As seen in the figure, satisfactory agreement is found. In Fig.~\ref{fig:rho_net}(b) we re-plot the same data as in panel (a), but on a double logarithmic scale. This allows us to focus on the early stages of the time evolution. As also appreciated in the inset, the average density of active links $\avg{\rho}$ quickly decays to a value of $\xi=0.6$, in-line with the prediction of Eq.~(\ref{eq:xi}).

\begin{figure}
\centering
\includegraphics[width=1\columnwidth]{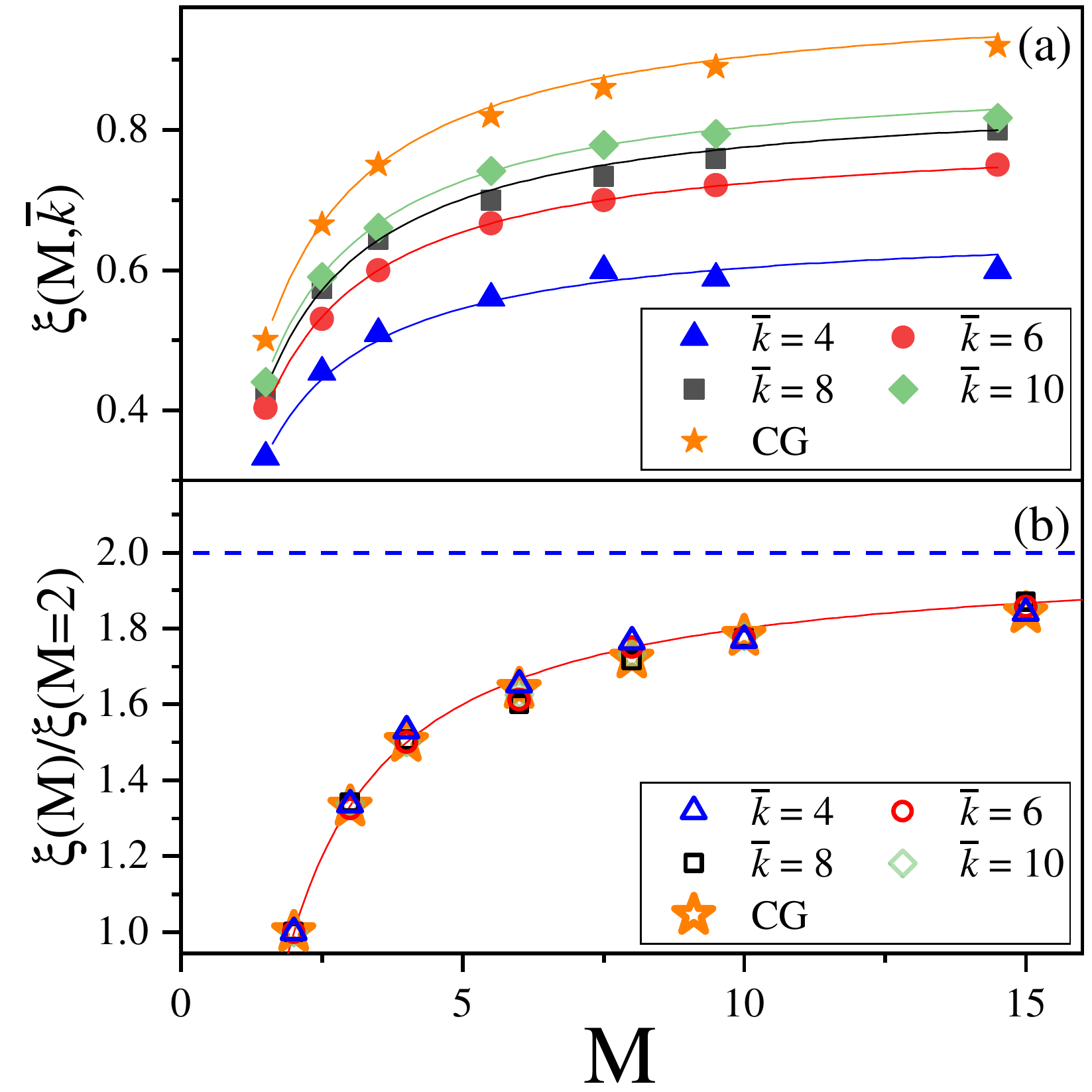}
\caption{(a) Density of active links, $\xi(M,\overline k$), in the meta-stable state as function of $M$ for different values of $\overline k$. Markers show simulations with homogeneous initial conditions for the complete graph (CG, orange stars), and for Erd\"os--Renyi network with different mean degrees (the model on the  Barab\'asi--Albert network has the same plateaux values as on Erd\"os--Renyi graphs). This data is obtained by measuring $\avg{\rho(t)}$, and then fitting to an exponential decay.  Lines are from Eq.~(\ref{eq:xi}), shown also for non-integer values of $M$ for optical convenience. Panel (b) shows $\xi(M,\overline k)/\xi(2,\overline k)$ as function of $M$. Symbols are the simulations from panel (a), the solid line is Eq.~(\ref{eq:scaling}). \label{fig:xi} }
\end{figure}

This initial decay of the density of active links is a signature of a rapid growth of local clusters of individuals with the same opinion state on the graph. As in the binary VM \cite{vazquez_eguiluz}, this initial decay is not described by the exponential law for $\avg{\rho(t)}$ from the pair approximation. After this initial phase the system is in a partially ordered metastable state, characterised by a density of active links $\xi(M,\overline k)$. The system  remains in this state indefinitely in the limit of infinite system size, we note that $\tau\to\infty$ for $N\to\infty$ in Eq.~(\ref{eq:tau_network}). If there are finitely many individuals in the network, then the system will eventually exit this state, triggered by fluctuations. Further ordering then occurs on a time scale of $\tau$, and the average density of active links decays exponentially. 

We now make some further observations about the partially ordered state. As seen in Fig.~\ref{fig:xi}(a), the density of active links in this initial metastable state, $\xi(M,\overline k)$, increases with the mean degree of the network, $\overline{k}$, and with the number of opinion states, $M$. From Eq.~(\ref{eq:xi}) we find
\begin{equation}\label{eq:scaling}
\frac{\xi(M, \overline k)}{\xi(2, \overline k)}= 2\left(1-\frac{1}{M}\right),
\end{equation}
as confirmed in Fig.~\ref{fig:xi}(b). We note the limiting value $\xi(M, \overline k )/\xi(2, \overline k) \to 2$ for $M\to\infty$.

\begin{figure}
\centering
\includegraphics[width=1\columnwidth]{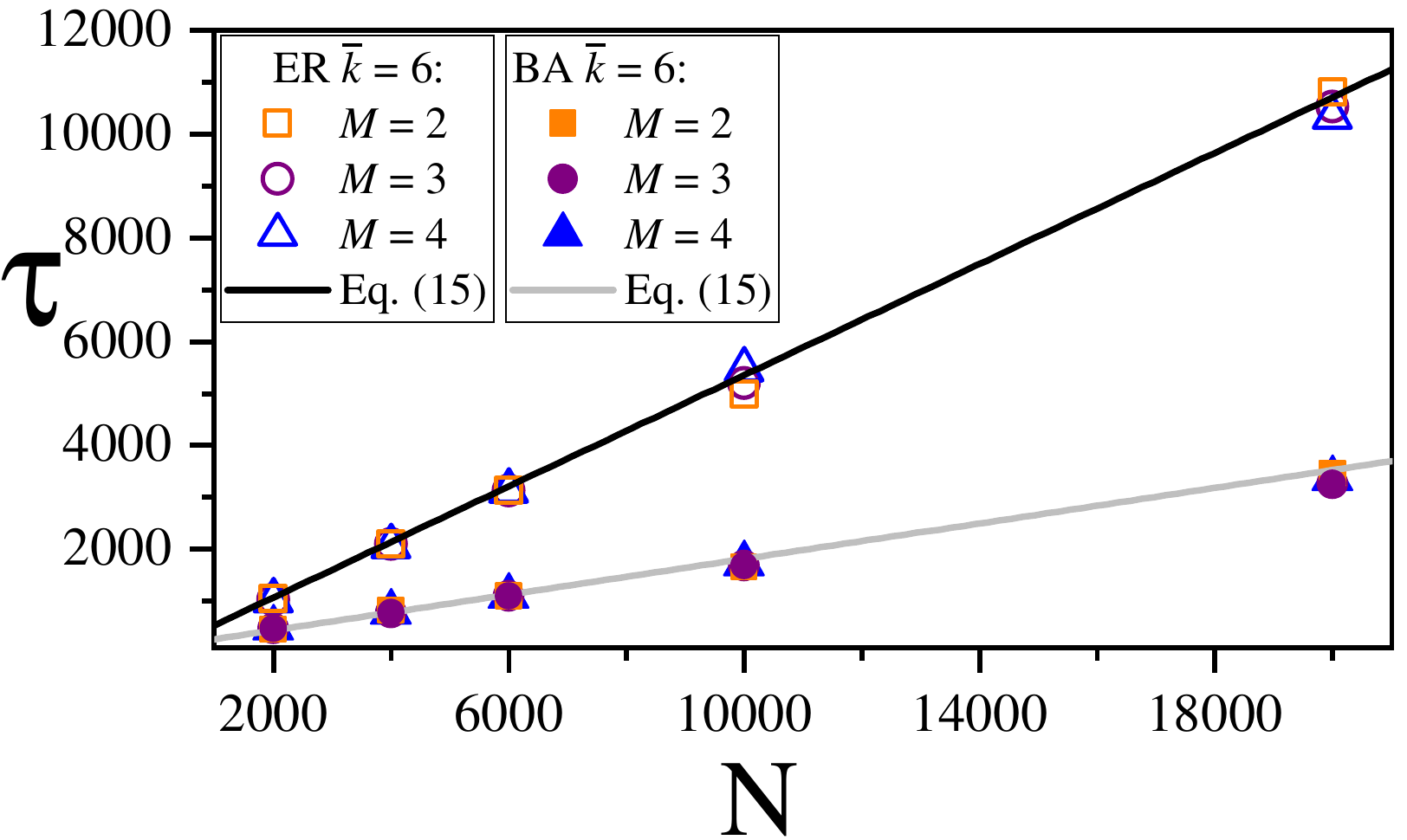}
\caption{Time scale $\tau$ of the exponential decay of the average density of active links for Erd\"os--Renyi (ER) graphs (empty symbols) and Barab\'asi--Albert (BA) networks (full symbols). The black and gray lines corresponds to Eq.~(\ref{eq:tau_network}) for ER ($\overline{k} = 6$) and BA ($\overline{k} = 6$), respectively. In-line with Eq.~(\ref{eq:tau_network}) $\tau$ is independent of $M$, i.e., the time scale $\tau$ in the MSVM is identical to that in the two-state model. Simulation data is obtained from measuring $\avg{\rho(t)}$, and a subsequent fit to an exponential decay. \label{fig:tau_network}}
\end{figure}

Finally, we briefly discuss the time scale $\tau$ in Eq.~(\ref{eq:rho_a_network}), which is given by the expression in Eq.~(\ref{eq:tau_network}), and does not depend on the number of opinion states $M$. As such the time scale in the multi-state model is the same as that in the conventional voter model with two opinion states. We note that the network structure enters not only through the mean degree $\overline k$, but also through the second moment $\overline{k^2}$ of the degree distribution. As a consequence the exponential decay of $\avg{\rho(t)}$ at fixed network size $N$ and mean degree is slower on an Erd\"os--Renyi graph than on a Barab\'asi--Albert network, see Figs.~\ref{fig:rho_net} and \ref{fig:tau_network}. As in the two-state model we have $\tau\propto N$ for complete graphs and ER networks, but $\tau\propto N/\ln\,N$ for large $N$ in the BA network \cite{vazquez_eguiluz, sood_redner, suchecki}.

\subsection{Time-evolution of average entropy}
\begin{figure}
\includegraphics[width=1\columnwidth]{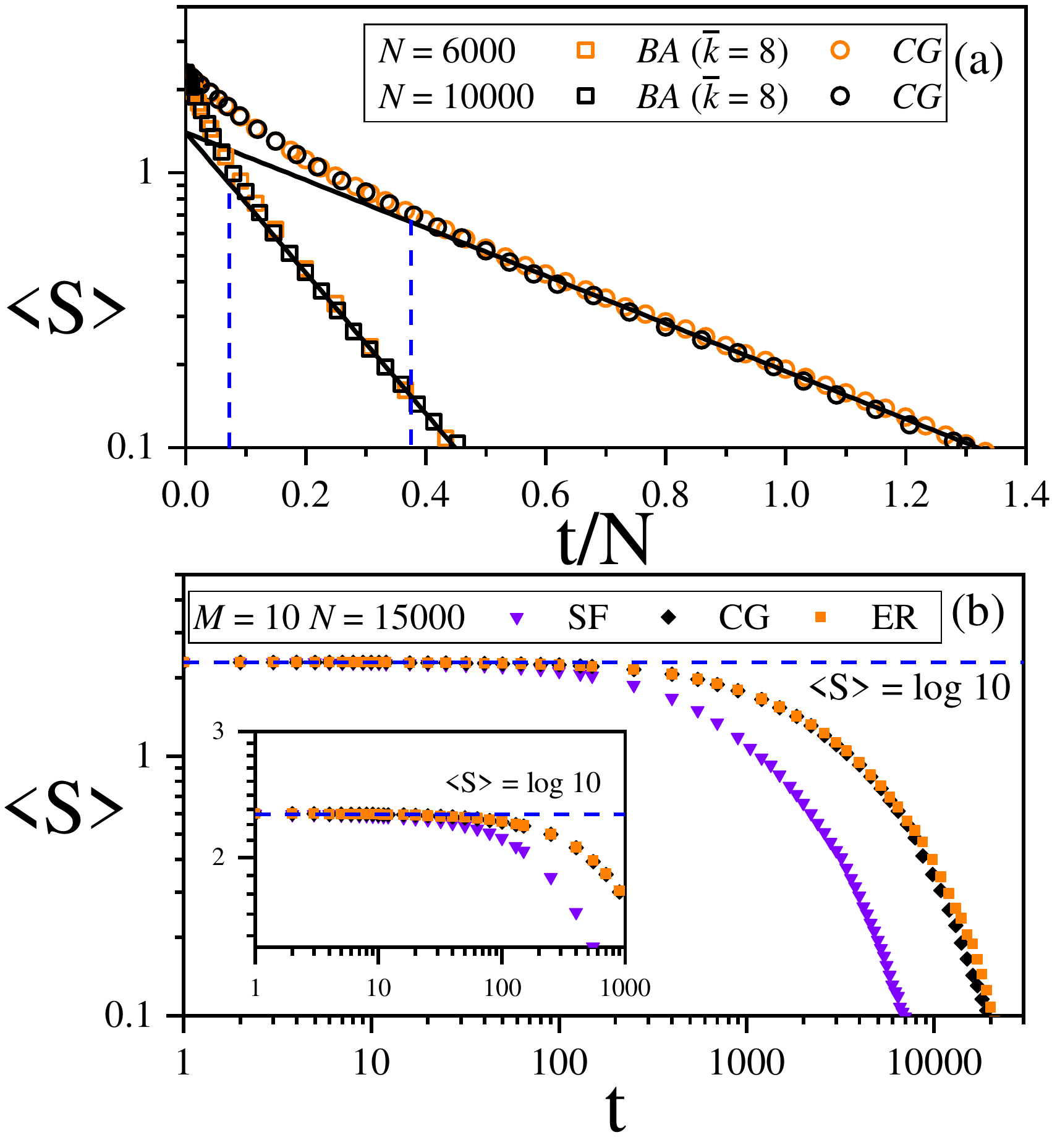}
\caption{Time evolution of the average entropy $\avg{S}$ for a system started from homogeneous initial conditions and $M = 10$. Panel (a) shows results on complete graphs (CG) and on Barab\'asi-Albert networks (BA, mean degree $\overline{k} = 8$) on linear-log scale. Markers are from simulations (averaged over $5000$ realisations). Lines are Eq.~(\ref{eq:S}) with $\tau$ given by Eq.~(\ref{eq:tau_all_to_all}) for the complete graph, and Eq.~(\ref{eq:tau_network}) for the BA networks. Vertical dashed lines indicate the time $t_2$ beyond which there are typically at most two opinions present in the population (see Sec.~\ref{sec:entr_calc}).  Panel (b) shows $\avg{S}$ on a doubly logarithmic scale for the CG (black diamonds), BA networks ($\overline{k} = 8$, purple triangles), and Erd\"os-Renyi graphs ($\overline{k} = 8$, orange squares), all for $M = 10$ and system size $N = 15000$. The horizontal dashed line is $S=S(t=0)=\ln\,M$. The inset shows a further zoom-in, and highlights that on networks there is no short-time drop of the entropy from the initial value. 
\label{fig:evolution_entropy}}
\end{figure}

\begin{figure}
\centering
\includegraphics[width=0.9\columnwidth]{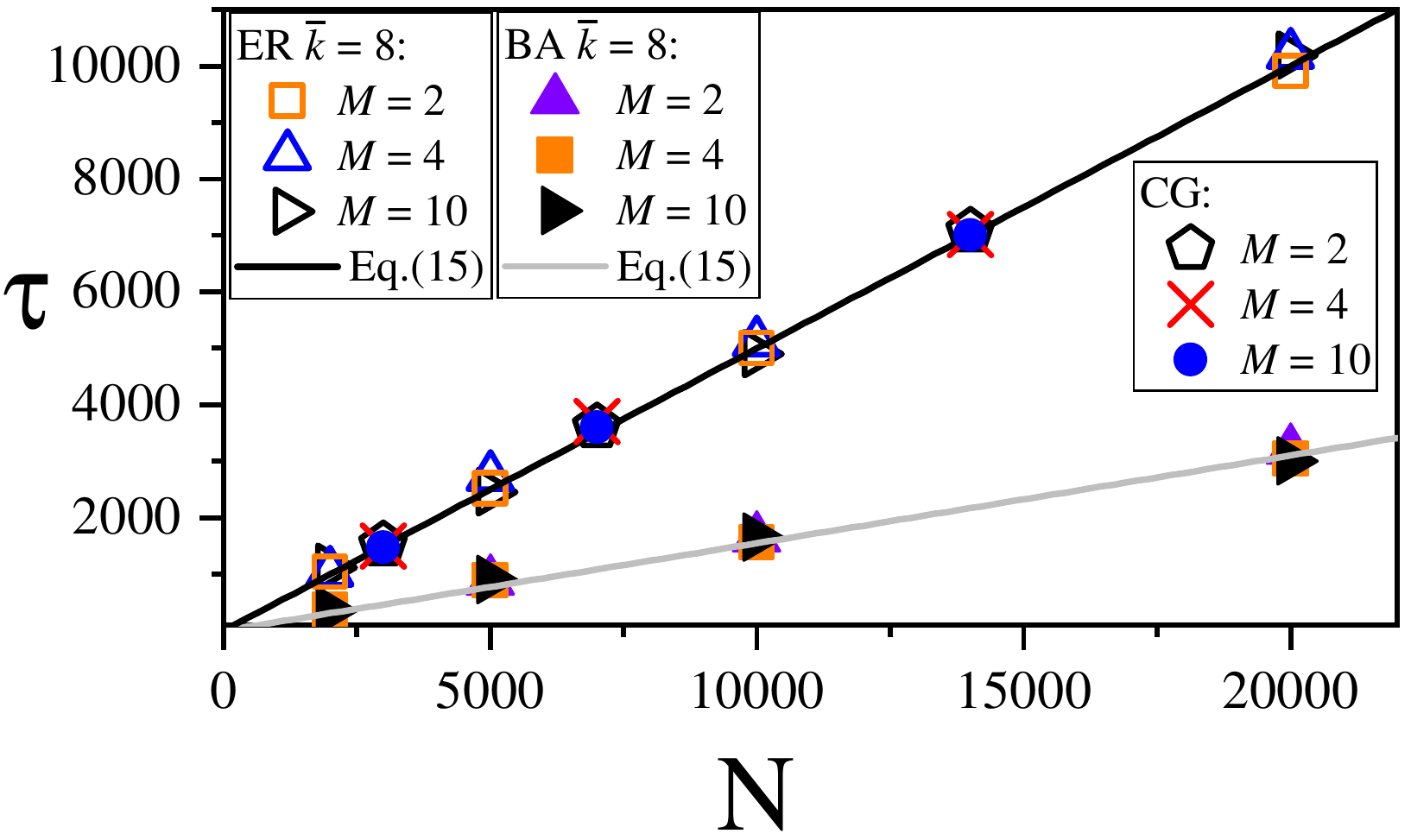}
\caption{Decay time of the average entropy for the complete graph (CG), and for Erd\"os--Renyi graphs (ER) and Barab\'asi--Albert networks (BA) with mean degree $\overline k = 8$ when, in average, there are two opinion states left. Markers are from fits of simulation data for $\avg{S(t)}$ to an exponential at large times. The black solid line correspond to Eq.~(\ref{eq:tau_network}), evaluated for the ER graph where hence $\tau \approx  N /2$  for $\overline k=8$, identical to the result for the complete graph.  The gray line is from Eq.~(\ref{eq:tau_network}) evaluated for BA graphs with $\overline{k} = 8$. 
 \label{fig:tau_S_network}}
\end{figure}

In numerical simulations we have also investigated the behaviour of the average entropy, $\avg{S(t)}=-\sum_\alpha\avg{x_\alpha \ln\,x_\alpha}$ over time. Simulation results are shown in Fig.~\ref{fig:evolution_entropy}. 
At long times the data is consistent with an exponential decay of the form
\be
\avg{S(t)}= \xi_S(M,N)\,e^{-t/\tau},~\mbox{for}~ t\gtrsim t_2,\label{eq:S}
\ee
see Fig.~\ref{fig:evolution_entropy}(a) and Fig.~\ref{fig:tau_S_network}. The decay time scale $\tau$ is the same as that for the average density of active links [e.g. Eq.~(\ref{eq:rho_a_network})]. The time $t_2$ from which (approximately) the decay of the mean entropy becomes exponential can be estimated as the mean time at which only at most two opinions are left in the population. This time can be approximated analytically (see  Sec.~\ref{sec:entr_calc} below).  When $M = 10$,  we find $t_2 \approx 0.38 N$ for complete graphs, and $t _2\approx 0.07 N$ on Barab\'asi--Albert networks.  These time points are indicated in Fig.~\ref{fig:evolution_entropy}(a). We stress that the deviations from an exponential for $t\lesssim t_2$ are not short-lived as those for the average density of active links on networks. Instead we find that $t_2$ scales linearly with the population size $N$.

Focusing on short times (of order $N^0$) in Fig.~\ref{fig:evolution_entropy}(b) we note the absence of an initial drop of the mean entropy on networks. Instead the mean entropy remains at its initial value for homogeneous initial conditions, $S(0)=\ln M$. This is in contrast with the behaviour of the average density of active links,  compare with Fig.~\ref{fig:rho_net}(b). 

This suggests the following picture of the coarsening process on networks:

(1) Once released from the homogeneous initial condition the system first undergoes a quick local relaxation process during the time interval up to $t=t^\star={\cal O}(N^0)$. In this phase the (average) density of active links reduces from its initial value $1-M^{-1}$ to the value $\xi(M,\overline k)$ given in Eq.~(\ref{eq:xi}). The average entropy  however, remains unchanged, $\avg{S(t)}=S(0)=\ln\,M$. This indicates that during this phase some local ordering takes place on the network (hence the reduction in $\avg{\rho}$), but that the proportions of individuals in the different opinion states do not materially change across the graph as a whole. One possible explanation is the formation of local domains. In each domain a particular opinion starts to outnumber the other opinion states. However, different domains do not `communicate', hence the effects of the local ordering average out across the system.

(2) After this initial phase the system tends towards consensus, and $\avg{\rho(t)}$ decays exponentially, with a time scale $\tau={\cal O}(N)$. The decay of the average entropy is not exponential until time $t\approx t_2={\cal O}(N)$.

(3) At long times ($t\gtrsim t_2$), when only two opinions are left in the system, the average entropy $\avg{S(t)}$ also decays exponentially, on the same time scale $\tau$ as the average density of active links.

We will discuss this further in Sec.~\ref{sec:entr_calc}.

\section{Path to consensus in individual trajectories} \label{sec:individual_realisations}

We now ask how representative the average behaviour discussed in Sec.~\ref{sec:ensemble} is of the dynamics of individual realisations. We first focus on the density of active links, and subsequently study the entropy of the distribution of individuals across the different opinion states.

\subsection{Evolution of the density of active links for individual realisations} \label{section:plateaux}

\subsubsection{Sequence of plateaux in individual realisations}

\begin{figure}
\includegraphics[width=0.8\columnwidth]{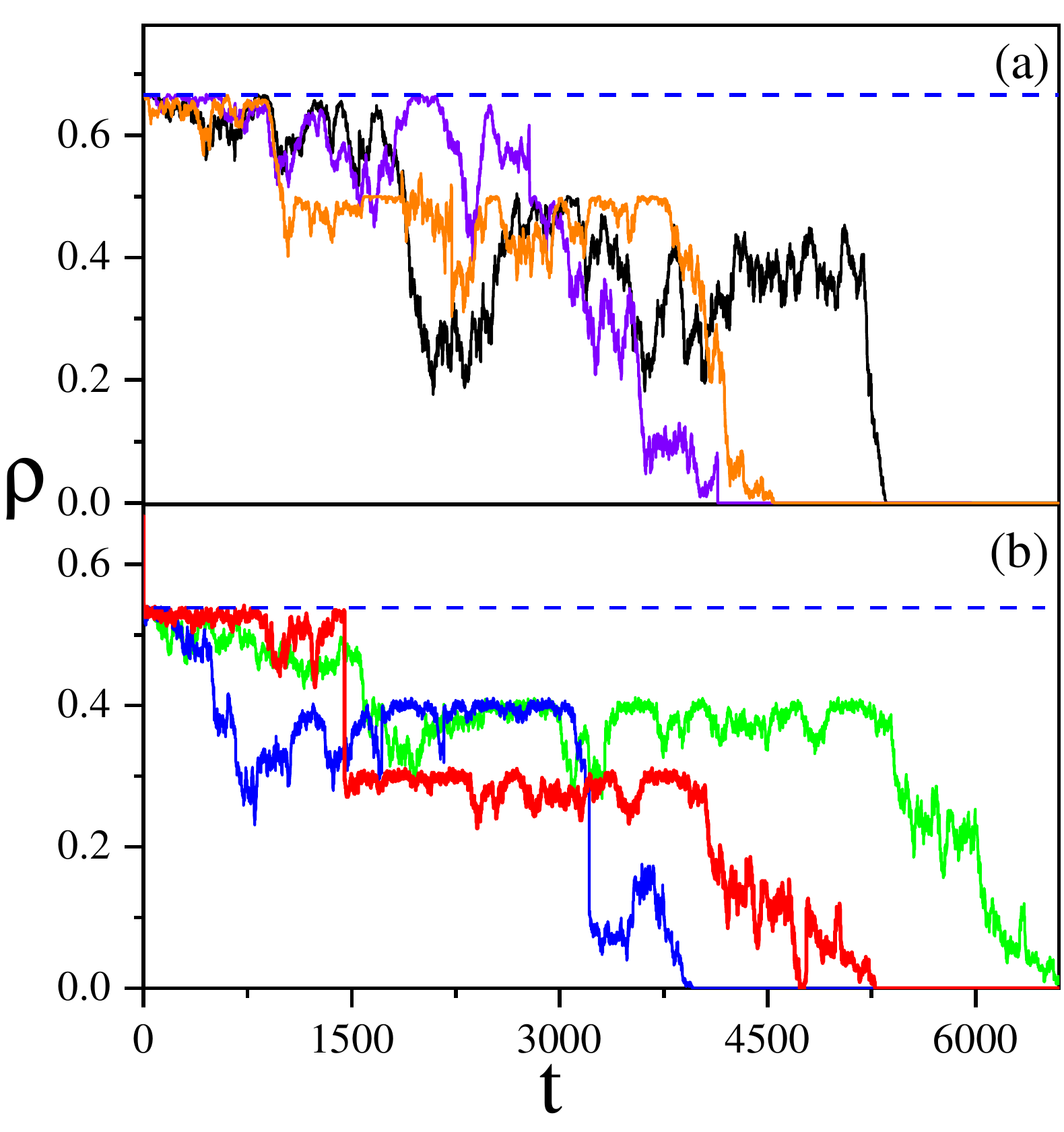}
\caption{Time evolution of the density of active links for  three individual realizations indicated by different colors. Panel (a) is on a complete graph ($M=3, N=6000$), panel (b) for an Erd\"os--Renyi graph ($\overline{k} = 6$, $M = 3$, $N = 6000$). Simulations were started from homogeneous initial conditions. The dashed lines indicate the values obtained from  Eq.~(\ref{eq:rho_hom}) [$\avg{\rho(0)} = 0.66$] and Eq.~(\ref{eq:xi}) [$\xi(M = 3, \overline{k} = 6) = 0.66$], for the complete graph and Erd\"os--Renyi graphs, respectively.   \label{fig:single}}
\end{figure}

Fig.~\ref{fig:single} illustrates the time evolution of the density of active links for individual realisations of the model with $M=3$ opinion states on a complete graph [panel (a)], and on an Erd\"os--Renyi graph [panel (b)]. 

The density of links is first found to fluctuate around an initial plateau (on networks this is preceded by a short transient.) For the complete graph this plateau is located at a point consistent with the initial value of the density of active links for homogeneous initial conditions,  $\rho = 1-M^{-1}$ [see Eq.~(\ref{eq:rho_hom})]. For the Erd\"os--Renyi network the density at this plateau is consistent with that predicted by Eq.~(\ref{eq:xi}).

Subsequently, the density of active links falls to a second plateau, where it then spends some time before a finite-size fluctuation takes the system to one of the absorbing states, where $\rho=0$. 

\begin{figure}
\centering
\includegraphics[width=0.8\columnwidth]{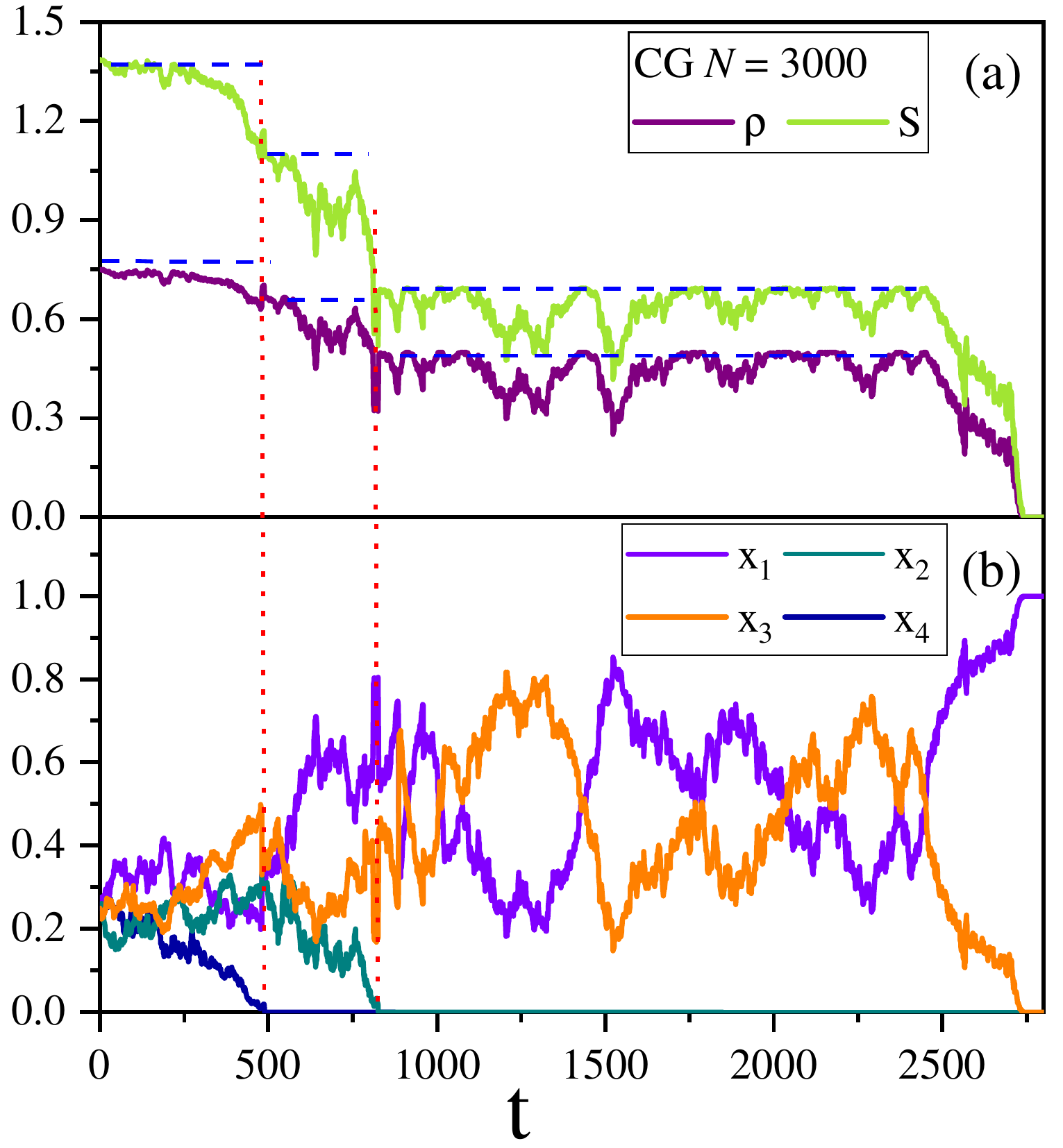}
\caption{Panel (a): Evolution of the density of active links (purple curve), of the entropy (green curve) for the model on a complete graph. Panel (b) shows the fraction of agents in each opinion state. Data is from a single simulation with $M = 4$ and $N = 3000$. The vertical dotted lines indicate the points in time at which the first and second opinions go extinct respectively. Dashed lines represent the plateau value $\xi$ considering a voter model with `$4$', `$3$', and `$2$' opinion states and the corresponding initial conditions; for the second and third plateau, the value was computed considering the number of agents in the surviving opinions once the extinction of opinions `$4$' and `$2$' took place.}\label{fig:individualCG}
\end{figure}


\begin{figure}[h]
\centering
\includegraphics[width=0.8\columnwidth]{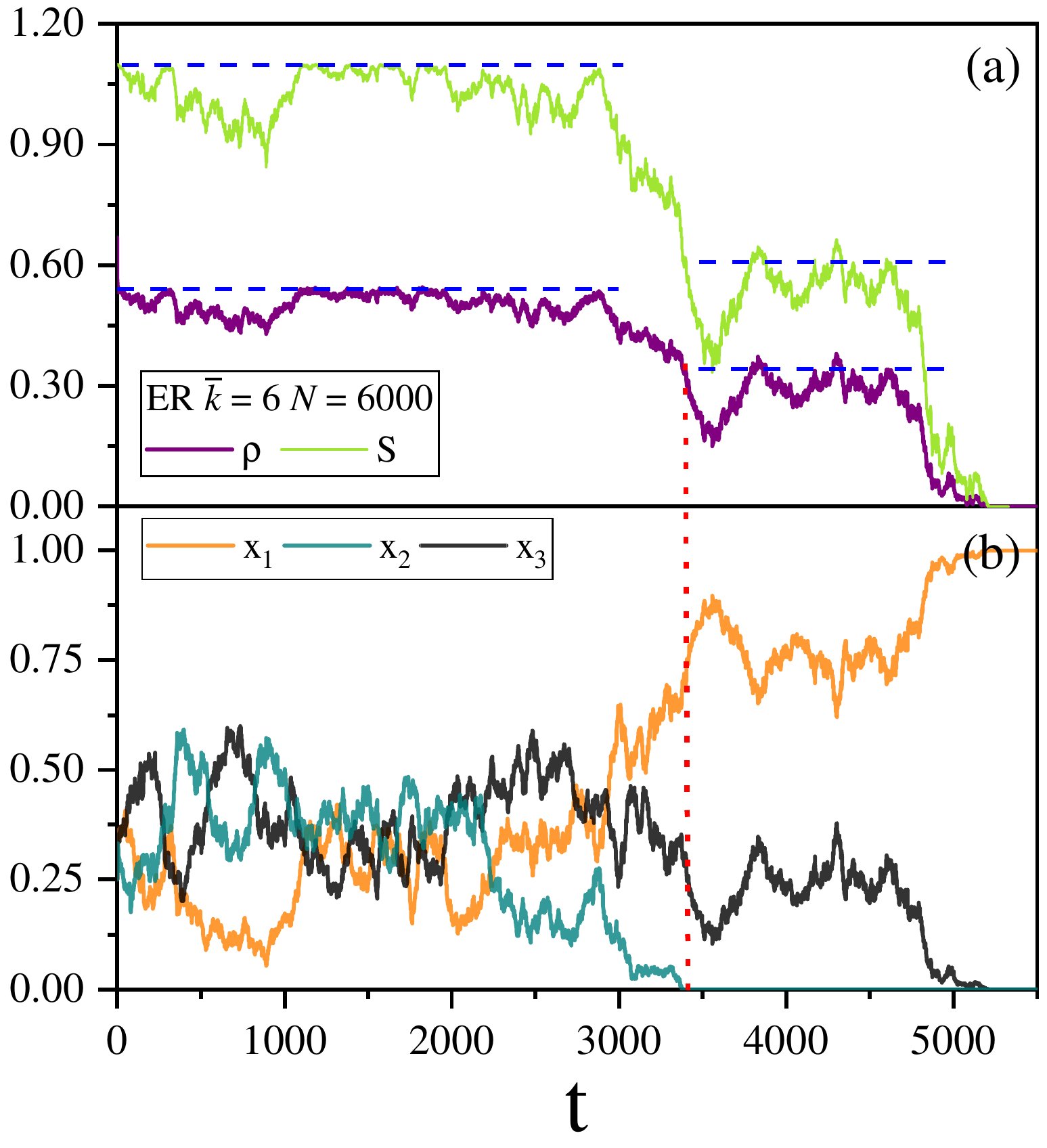}
\caption{Panel (a): Evolution of the density of active links (purple curve) and of the entropy (green curve) for the model on Erd\"os-Renyi graphs. Panel (b) shows the fraction of agents in the different opinion states. Data is from  a single simulation with $M = 3$ and $N = 6000$. The graph has mean degree $\overline k=6$.  The vertical dotted line indicates the time at which the first opinion state becomes extinct. Dashed lines represent the plateau value $\xi$ considering a voter model with  `$3$' and `$2$' opinion states and the corresponding initial conditions; the second plateau value was obtained considering the number of agents in the surviving opinions once opinion state  `$2$'  disappears. \label{fig:individualER} }
\end{figure}

Unlike the initial plateau, the value of the density of active links at this second plateau differs from realisation to realisation (see Fig.~\ref{fig:single}). The process leading to this intermediate plateau can be better understood from the inspection of the evolution of the number of agents in each opinion state in Figs.~\ref{fig:individualCG}(b) and \ref{fig:individualER}(b).  As shown, opinions go extinct one after the other. Each extinction takes the density of active links $\rho$ to a new plateau (at a value lower than that of the previous plateau), until consensus is reached and $\rho = 0$. 

When $K$ extinctions have taken place in the model with initially $M$ opinions, we are left with a MSVM with $L=M-K$ states. Crucially however, the initial condition for this model with $L$ states is the result of the previous dynamics up to the time when the $K$-th extinction takes place. This initial condition in turn determines the value of the density of active links at the subsequent plateau. Given that different realisations result in different configurations at the time of the extinctions the ensuing plateau densities also vary across realisations. This is the reason for the spread of plateaux in Fig.~\ref{fig:single}.  

\subsubsection{Density of active links at the different plateaux}

We now proceed to calculate the mean density of active links in the model with initially $M$ opinions at the first point in time where only $L$ opinions are left. Suppressing any possible dependence on $M$, we will denote this quantity by $\avg{\rho}_L$.  

 Without loss of generality we assume that the $L$ opinions left in the population are  $\alpha=1,\dots,L$. For an all-to-all interaction We then have
\be
\rho=\frac{1}{2}\sum_{\alpha=1}^L [2x_\alpha (1-x_\alpha)],
\ee
where $\sum_{\alpha=1}^L x_\alpha=1$. Each term $2x_\alpha (1-x_\alpha)$ accounts for links involving individuals of type $\alpha$ connected with individuals of any other opinion, and the overall pre-factor $1/2$ corrects for double counting. After taking an average over realisations we thus have
\BE\label{eq:rho_L_aux}
\avg{\rho}_L&=&\avg{\sum_{\alpha=1}^L x_\alpha (1-x_\alpha)} \nonumber \\
&=&L \int_0^1 dx \,P_L(x) x(1-x),
\EE
where the quantity $P_L(x)$ is the distribution of the fraction of agents found in a particular opinion state when only $L$ opinions are left. We have used the fact that, by symmetry, no opinion state is preferred over any other.

The distribution $P_L(x)$ can be obtained making the following hypothesis. We assume that, at the extinction point leaving only $L$ opinions in the system, all configurations with $x_\alpha\geq 0$ and $\sum_{\alpha=1}^L x_\alpha=1$ are equally likely, i.e., the distribution of $(x_1,\dots,x_L)$ at that time is assumed to be
\be
P_L(x_1,\dots,x_L)=[(L-1)!]\times\delta\left(\sum_{\alpha=1}^L x_\alpha-1\right),
\ee
where $\delta(\cdot)$ is the Dirac delta function. 

The distribution $P_L(x)$ in Eq.~(\ref{eq:rho_L_aux}) is the single-variable marginal of $P_L(x_1,\dots,x_L)$,
\BE\label{eq:plx_main}
P_L(x_1)&=&(L-1)!\int dx_2\cdots dx_L \,\delta\left(\sum_{\alpha=1}^L x_\alpha-1\right) \nonumber \\
&=& (L-1)(1-x_1)^{L-2}.
\EE
For further details see Appendix~\ref{app:simplex}.

We can then directly calculate $\avg{\rho}_L$ in Eq.~(\ref{eq:rho_L_aux}), 
\be\label{eq:rho_L}
\avg{\rho}_L=\frac{L-1}{L+1}.
\ee
The argument so far applies to complete graphs, where the initial condition is known to directly set the typical density of active links at the plateau that follows (see Sec.~\ref{sec:ensemble}). 
As also discussed in Sec.~\ref{sec:ensemble}, the system undergoes a brief transient when it is started on an uncorrelated graph, and the subsequent plateau value of $\avg{\rho}$ is obtained applying a multiplication factor  $(\overline{k}-2)/(\overline{k}-1)$. For uncorrelated graphs we therefore predict
\be \label{eq:rho_L_graph}
\avg{\rho}_L= \frac{\overline{k}-2}{\overline{k}-1}\frac{L-1}{L+1} .
\ee

\subsubsection{Test against simulations}
\begin{figure}
\includegraphics[width=1\columnwidth]{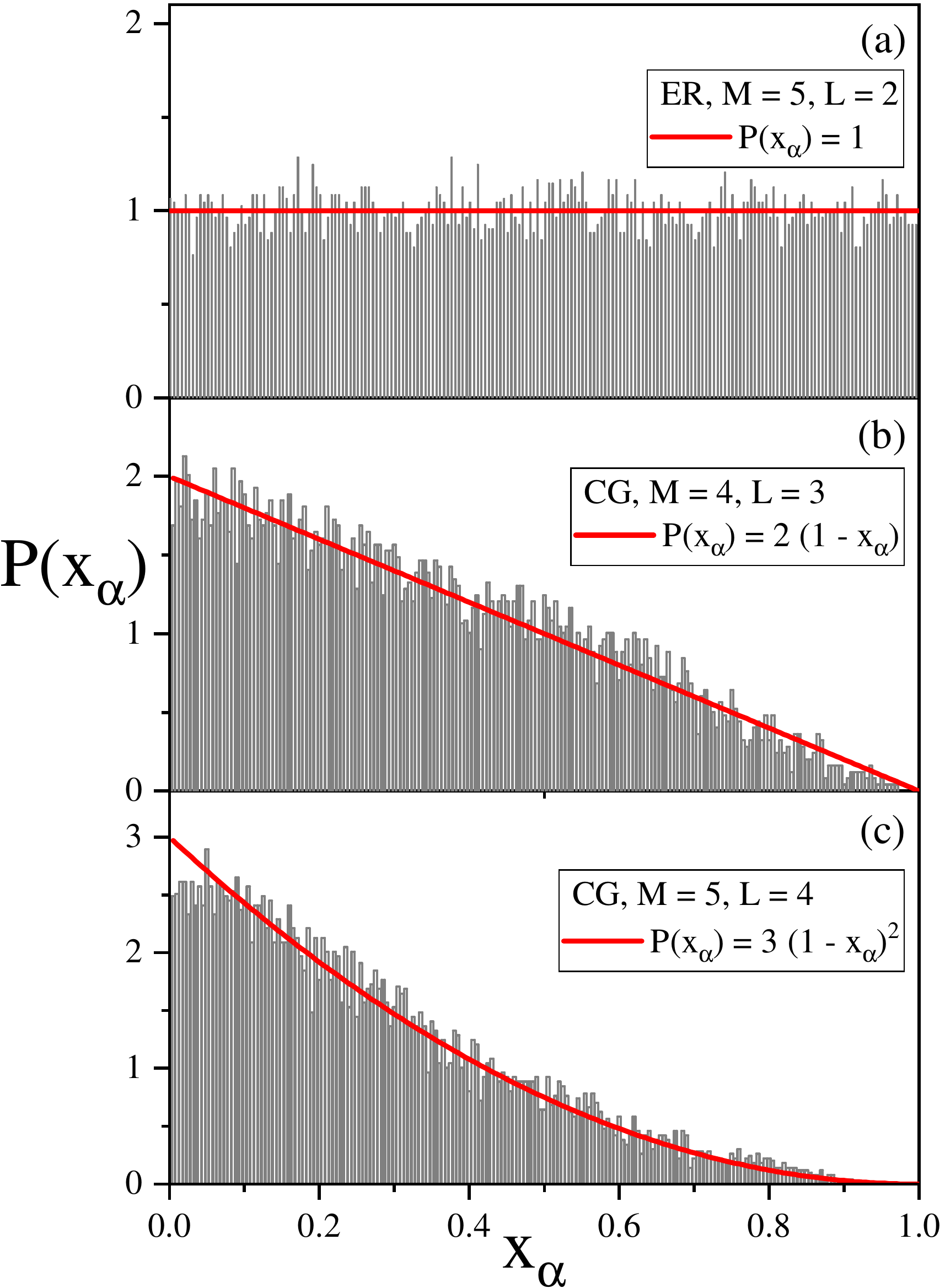}
\caption{Marginal distribution $P_L(x_\alpha)$ for the fraction of agents in any one opinion at the first point in time when $L$ opinions are left in a model with initially $M>L$ states. Gray bars are from numerical simulations ($N=5000$, averaged over $5000$ realisations), the solid lines are Eq.~(\ref{eq:plx_main}). Panel (a) is for Erd\"os-Renyi networks, panels (b) and (c) for complete graphs.\label{fig:histograma}}
\end{figure}

We now test these predictions against simulations. First, we verify the validity of our hypothesis of a flat distribution for $(x_1,\dots,x_L)$ at the first point in time when there are only $L$ opinions in the population. Fig.~\ref{fig:histograma} shows simulation results for the marginal distribution for $x_\alpha$ for different choices of $M$ and $L$ on complete graphs and on Erd\"os--Renyi networks. As can be seen from the figure, these simulations are consistent with the predictions of Eq.~(\ref{eq:plx_main}).

\begin{figure}
\centering
\includegraphics[width=1\columnwidth]{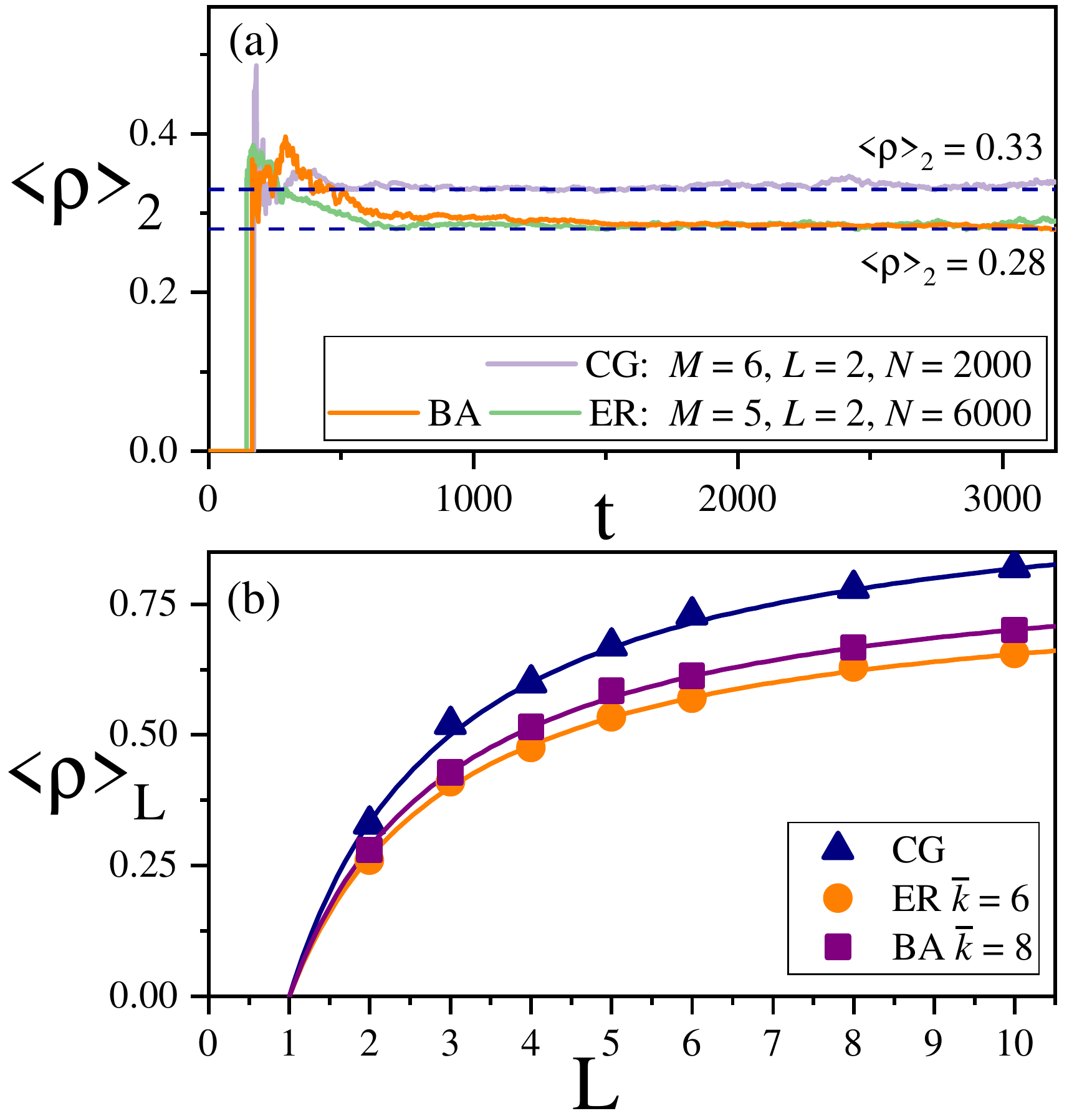}
\caption{(a) Evolution of the density of links in realisations for which only $L = 2$ opinion states are left in the system for the CG, and the ER and BA networks with $\overline{k} = 8$. The plateau is located at $\avg{\rho}_{L=2} \approx 0.33$ for the CG and at $\avg{\rho}_{L=2} \approx 0.28$ for the networks, in good agreement with Eq.~(\ref{eq:rho_L})  and Eq.~(\ref{eq:rho_L_graph}), respectively. Panel (b) shows the location of the intermediate plateaux, $\avg{\rho}_L$ as a function of $L$ in a model with initially $M=15$ states. Markers are from simulations on CG (triangles), on BA networks (squares), and on ER networks (circles). Lines are from Eqs.~(\ref{eq:rho_L}) and (\ref{eq:rho_L_graph}) respectively.   \label{fig:int_plateau}}
\end{figure}

We next introduce the concept of restricted ensemble at a given time. This is the ensemble of trajectories that, at this time, have precisely $L$ surviving opinions. In Fig.~\ref{fig:int_plateau} we show the average of the density of active links over this restricted ensemble. The data in panel (b) was obtained by first averaging over the restricted ensemble at any given time, and subsequently an average over time is performed. Fig.~\ref{fig:int_plateau}(b) thus demonstrates that the average density of interfaces at the first point in time at which there are only $L$ opinions in the system is given by Eqs.~(\ref{eq:rho_L}) and (\ref{eq:rho_L_graph}) for all-to-all interaction and on networks respectively, and that this average density of active interfaces is then maintained by the system until the next extinction occurs.

\subsection{Connection to exponential decay of the ensemble averaged density of links}\label{sec:connect}
\subsubsection{Sequence of `jumps' in the model with multiple opinions}
We have described the dynamics at the ensemble level and at the level of individual realisations. We now proceed to a characterization in terms of the restricted ensembles that we have just introduced.

The average density of active links, $\avg{\rho}$ decays exponentially, as indicated in Eqs.~(\ref{eq:exp}) for complete graphs, and in Eq.~(\ref{eq:rho_network}) for uncorrelated networks. The decay time scales are given in Eqs.~(\ref{eq:tau_all_to_all}) and (\ref{eq:tau_network}) respectively.
\begin{figure}
\centering
\includegraphics[width=0.85\columnwidth]{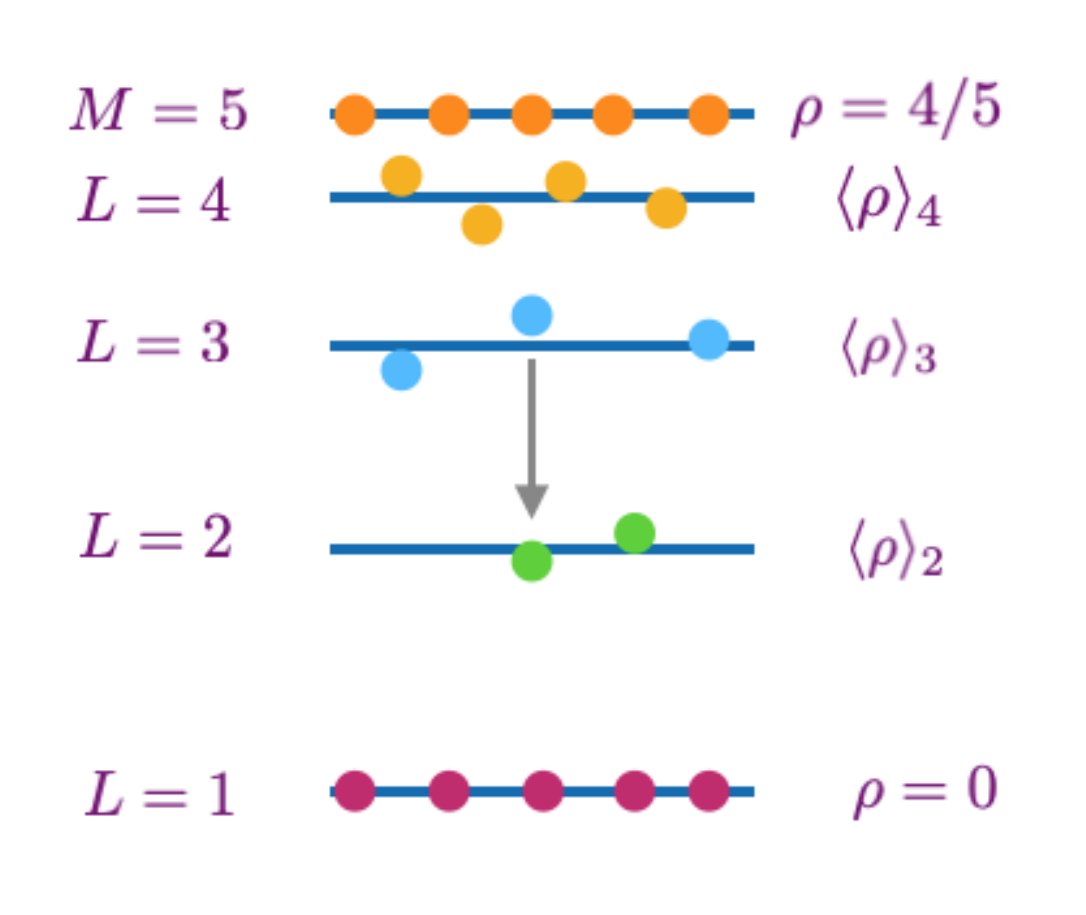}
\caption{Illustration of the jump process different realisations of the MSVM undergo. Each jump is associated with the extinction of one opinion. The density of active links in the resulting sequence of plateaux differs from realisation to realisation (as indicated by the scatter of markers), the mean density in the plateaux is given by Eqs.~(\ref{eq:rho_L}) or (\ref{eq:rho_L_graph}) respectively. In the final state ($L=1$) consensus has been reached, and hence $\rho=0$. The residence time in each level varies from realisation to realisation as well, the mean time spent in level $L$ is given by Eq.~(\ref{eq:delta_t}). The illustration is for a model with initially $M=5$ states. For homogeneous initial conditions one then has $\rho=4/5$. \label{fig:levels}}
\end{figure}

Individual realisations can be characterised as undergoing a sequence of `jumps' in the density of active links, from one plateau to another, as illustrated in Fig.~\ref{fig:levels}. Each of these jumps is associated with the extinction of an opinion. As we have seen the density of active links along the sequence of plateaux differs across realisations. We have established that the average density of active links at the plateau at which $L$ opinions are left in the population is given by Eq.~(\ref{eq:rho_L}) for complete graphs, and by Eq.~(\ref{eq:rho_L_graph}) on uncorrelated networks.

\subsubsection{Exponential decay in the ensemble on complete graphs}
We would now like to connect the observations at ensemble level with those at the level of realisations. In order to do this we need information about the typical residence time in the different plateaux. We first discuss this for the model on the complete graph.

Following \cite{starnini} the mean time the system spends in the metastable state with $L$ opinions can be estimated as the difference between the mean consensus times in voter models with initially $L$ or $L-1$ states respectively. The consensus time from a state with $K$ opinions in the model with all-to-all interaction is given by \cite{starnini}
\be
\avg{T_N(K)}=N\frac{K-1}{K},
\ee
for a system of size $N$, and assuming that all initial conditions are equally likely. We then have the following mean residence time in the state with exactly $L$ opinions,
\BE\label{eq:delta_t}
\avg{\Delta t}_L &=&  \avg{T_N(L)} - \avg{T_N(L-1)} \nonumber \\ &=& \frac{N}{L(L-1)}.
\EE
We also know that the mean density of active links at the plateau with $L$ opinions is $\avg{\rho_L}=(L-1)/(L+1)$ [Eq.~(\ref{eq:rho_L})]. Therefore the mean change of the density of links when transitioning from the plateau with $L$ opinions to that with $L-1$ opinions is
\BE
\avg{\Delta\rho}_{L\to L-1}&=&
\avg{\rho}_{L-1}-\avg{\rho}_L\nonumber \\
&=& \frac{-2}{L(L+1)}.
\EE
Using these results for $\avg{\Delta\rho}_{L\to L-1}$, $\avg{\Delta t}_L$ and $\avg{\rho}_L$ we conclude
\be\label{eq:aux1}
\frac{\avg{\Delta \rho}_{L\to L-1}}{\avg{\Delta t}_L}=-\frac{2}{N}\avg{\rho_L}.
\ee
The left hand side is a proxy for the time derivative of $\avg{\rho}$ when there are exactly $L$ opinions in the system. The mean density of active links in this situation is $\avg{\rho}_L$. Therefore we have $\frac{d}{dt}\avg{\rho}_L=-(2/N)\avg{\rho}_L$, and we recover the exponential decay in Eq.~(\ref{eq:exp}). The time scale of the decay matches that in Eq.~(\ref{eq:tau_all_to_all}), up to the replacement $N\to N-1$ (which becomes irrelevant for large $N$). 

The exponential decay law [Eq.~(\ref{eq:exp2})] for the average density of active links in the ensemble can therefore be recovered from the picture of jumps in Fig.~\ref{fig:levels}, and is a consequence of the specific relation between the level spacings and the mean residence time in each level.

\subsubsection{Uncorrelated networks}
A similar argument applies on uncorrelated networks. The $\avg{\rho}_L$, and hence the $\avg{\Delta\rho}_{L\to L-1}$, are then multiplied by the common factor $(\overline k-2)/(\overline k-1)$, see Eq.~(\ref{eq:rho_L_graph}). This therefore drops out in Eq.~(\ref{eq:aux1}). In order to recover the exponential decay in Eq.~(\ref{eq:rho_a_network}) with a decay time as in Eq.~(\ref{eq:tau_network}) we then need to show that the result of \cite{starnini} generalises to uncorrelated graphs as follows,
\be
\avg{T_N(K)}=N\frac{K-1}{K} \times\frac{(\overline k-1)\overline{k}^2}{(\overline{k}-2)\overline{k^2}}.\label{eq:TNK_network}
\ee

\begin{figure}
\centering
\includegraphics[width=1\columnwidth]{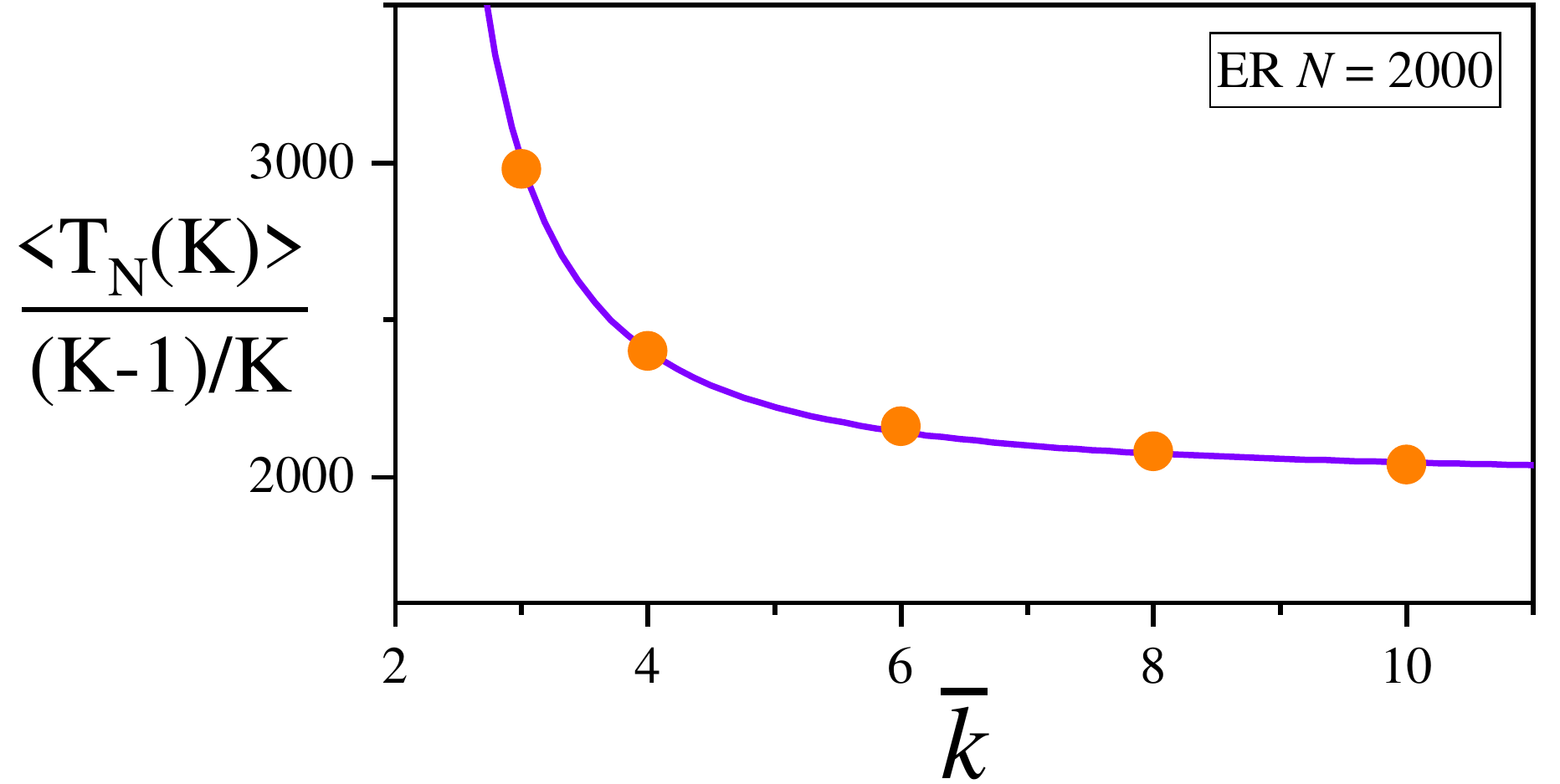}
\caption{Test of Eq.~(\ref{eq:TNK_network}) for the model on Erd\"os-Renyi networks with $N = 2000$. Markers represent the ratio $\avg{T_N(K)}/[(K-1)/K]$ from simulations. The line is $[\overline k-1)\overline{k}^2]/ [(\overline{k}-2)\overline{k^2}]$, where $\overline{k^2}=Np(1-p)$, with $p=\overline{k}/(N-1)$. \label{fig:TNK_network}}
\end{figure}

This can be demonstrated using a reduction to an effective two-state model, and the  properties of the two-state VM on uncorrelated graphs, where it is known that the relevant time scales undergo re-scaling by a factor $[(\overline k-1)\overline{k}^2]/[(\overline{k}-2)\overline{k^2}]$ relative to the case of all-to-all connectivity \cite{vazquez_eguiluz}. This is described in more detail in Appendix~\ref{app:ext}. The validity of Eq.~(\ref{eq:TNK_network}) is demonstrated in Fig.~\ref{fig:TNK_network}.

\subsection{Evolution of the entropy for individual realisations and in the ensemble average}\label{sec:entr_calc}
\subsubsection{Absence of simple exponential decay at ensemble level}
 
\begin{figure}
\centering
\includegraphics[width=0.9\columnwidth]{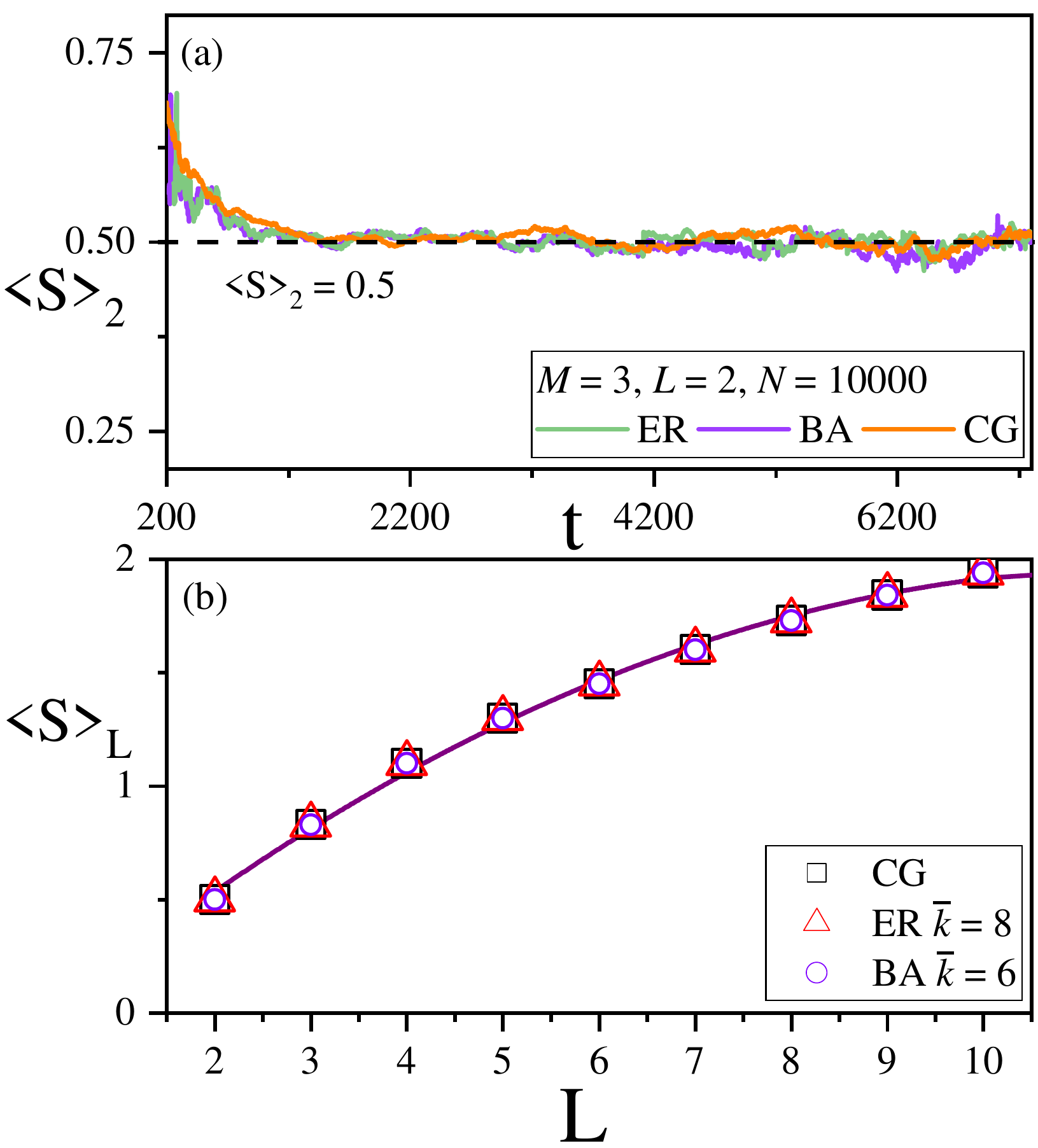}
\caption{(a) Time evolution of the entropy, at any time averages only over realisations for which exactly $L= 2$ opinions are present in the population. Simulations are for the CG, ER, and BA graphs of size $N = 10000$ and initially $M=3$ opinion states. The plateau is located at $\avg{S}_L= 0.5$ in agreement with Eq.~(\ref{eq:plateau_entropy}). (b) Symbols show simulation results for $\avg{S}_L$ as a function of $L$ in a model with initially $M=15$ opinions. Markers are from simulations for the CG (squares), ER graphs (triangles) and BA networks (circles). These are obtained from performing a time average on data such as the one in panel (a). The solid line is the analytical prediction in Eq.~(\ref{eq:plateau_entropy}). \label{fig:plateauS} }
\end{figure}

In Figs.~\ref{fig:individualCG} and \ref{fig:individualER} we also observe intermediate plateaux in the time evolution of entropy for individual realisations. In  Fig.~\ref{fig:plateauS}(a) we therefore proceed as we did for the density of active links and perform averages over restricted ensembles of realizations with fixed $L$  surviving opinions, $L<M$,  at different times. In the example in the figure, we find that $\avg{S}_{L=2}\approx 0.5$. 
 
Analytically, we find
\BE\label{eq:plateau_entropy}
\notag \avg{S}_L&=& - L \int_0^1 dx\, P_L(x)\, x \ln\,x \\
&=& H_L -1 ,
\EE
where the $H_L=\sum_{\ell=1}^L (1/\ell)$ are the harmonic numbers. The distribution $P_L(x)$ is given in Eq.~(\ref{eq:plx_main}). The prediction in Eq.~(\ref{eq:plateau_entropy}) is tested against numerical simulations in Fig.~\ref{fig:plateauS}(b). 

We can now use this to understand in more detail why the decay of the entropy at ensemble level does not follow an exponential law initially, but becomes exponential at long times. From Eq.~(\ref{eq:plateau_entropy}) we find $\avg{\Delta S}_{L\to L-1}=1/L$. Using Eq.~(\ref{eq:delta_t}) we then have
\be\label{eq:aux_S1}
\frac{\avg{\Delta S}_{L\to L-1}}{\avg{\Delta t}_L}=-\frac{L-1}{N(H_L-1)}\avg{S_L}.
\ee
Interpreting the left-hand side again as a time derivative, we therefore have 
\be\label{eq:entr_decay}
\frac{d}{dt} \avg{S}=-\frac{1}{\tau_S(L)} \avg{S}
\ee
in the time regime when there are $L$ opinions left in the system, with 
\be\label{eq:tau_S_L}
\tau_S(L)=N\frac{H_L-1}{L-1}.
\ee
for complete graphs, and $\tau_S(L)=N\frac{H_L-1}{L-1}\times
\frac{(\overline k-1)\overline{k}^2}{(\overline{k}-2)\overline{k^2}}$ on uncorrelated graphs.

The pre-factor $-1/\tau_S(L)$ on the right in Eq.~(\ref{eq:entr_decay}) explicitly depends on $L$. In the corresponding equation (\ref{eq:aux1}) for the average density of active links the pre-factor is instead constant. This is why the average density of links decays exponentially, and the average entropy does not. For $L=10$, $\tau_s(L)$ evaluates to (approximately) $0.21N$, for $L=9$ to $0.23N$, for $L=8$ to $0.25N$ and so on. This demonstrates that the relative decay of the average entropy $\avg{S(t)}$ slows down as more and more opinions become extinct.

We note that at any one time $t$ different realisations of the MSVM process will be in states with different numbers of opinions $L$ left in the population. It is therefore not straightforward to aggregate the mechanics of the decay of entropy for a fixed value of $L$ [Eq.~(\ref{eq:entr_decay})] into a global picture for the behaviour of entropy at the ensemble level. This is at variance with the decay of the average density of links. Given that all $\tau_s(L)$ scale linearly in $N$, we can however conclude that all timescales governing the decay of entropy are ${\cal O}(N)$, as also demonstrated in Fig.~\ref{fig:evolution_entropy}~(a).

\subsubsection{Regime at large times}
One can identify a regime in which most realisations will either have reached consensus or in which the population contains only two opinions. In a model with initially $M$ opinions this will be the case at times which are  greater than 
\be\label{eq:t_2}
t_2\equiv \sum_{L =3}^M \avg{\Delta t}_L,
\ee
This is the sum of average times spent in states with $M, M-1, \dots, 3$ opinion states respectively, i.e., it is the mean time until only two opinion states are left in the system.

When $L=2$ opinions are left, we find $\tau_S(L)=N/2$ for the complete graph in Eq.~(\ref{eq:tau_S_L}). In this long-time regime, we then have $\frac{d}{dt} \avg{S}_2=-(2/N)\avg{S}_2$, leading to an exponential decay with the same time scale as that for the average density of active links. On uncorrelated networks this time scale is multiplied by the factor $(\overline k-1)\overline{k}^2/[(\overline{k}-2)\overline{k^2}]$.

\begin{figure}
\centering
\includegraphics[width=0.9\columnwidth]{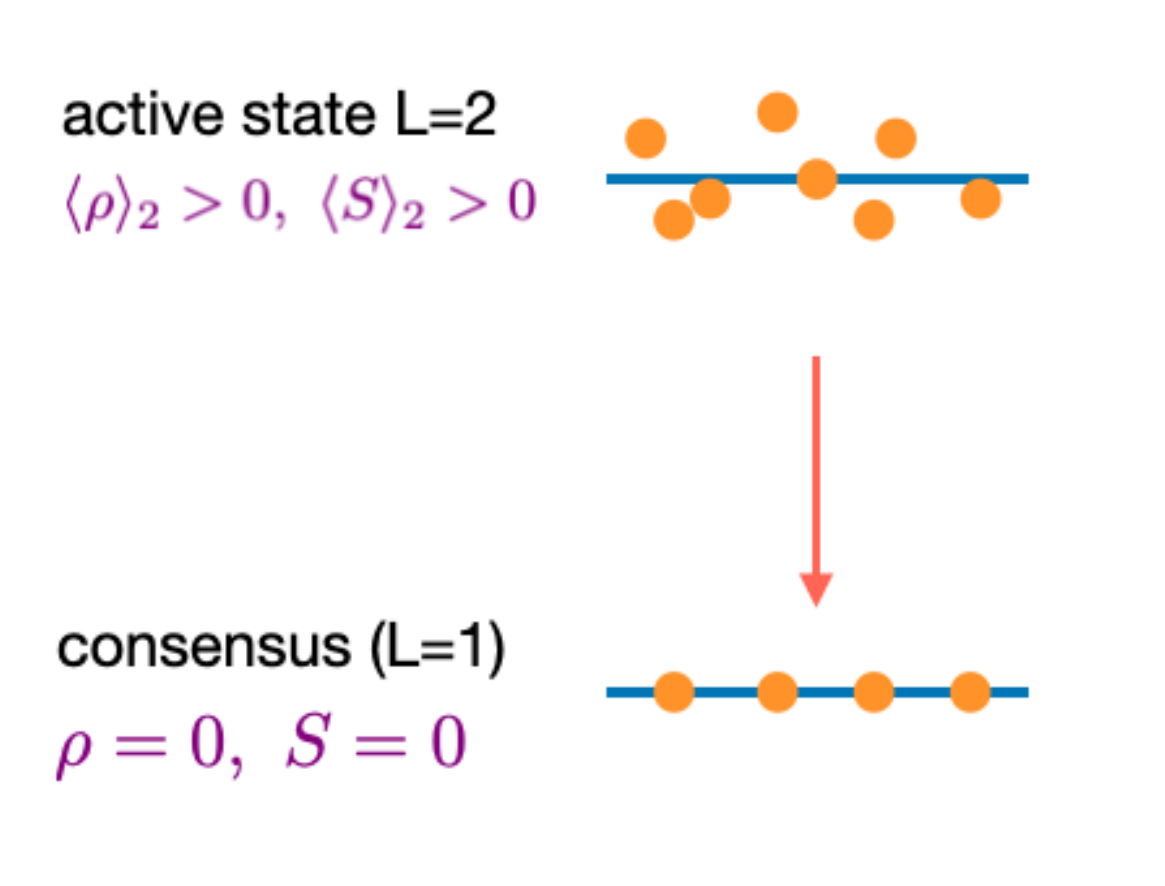}
\caption{Illustration of the dynamics when only at most two opinions are left in the population. Some realisations are in the active state (two opinions). Each such realisation is in a meta-stable state with different plateau values for $\rho$ and $S$ across realisations. Transitions to consensus occur with rate $\tau_S(L=2)^{-1}=2/N$, leading to exponential decay of $\avg{\rho}$ and $\avg{S}$ with a decay time scale $N/2$.\label{fig:2level} }
\end{figure}

We can interpret this as the dynamics of a two-level system, see Fig.~\ref{fig:2level} for an illustration. Each realisation will either have two opinions left in the population, or have reached consensus. There is hence an active state (two opinions) and a passive consensus state. Realisations in the active state have positive densities of active links and positive entropies. There is some scatter across realisations, averages among active realisations are given by $\avg{\rho}_2>0$ and  $\avg{S}_2>0$, respectively. In the consensus state both $\rho$ and $S$ are zero. Each realisation is first in the upper level ($L=2$), and then reaches consensus through a sudden fluctuation. This can be thought of as a Markovian jump process. Realisations transition from the upper to the lower level independently, with rate $2/N$. The population of the upper level hence decays exponentially in time. This then leads to the exponential decay of both $\avg{\rho}$ and $\avg{S}$. 

\section{Stabilising intermediate states through the introduction of zealots} \label{sec:zealots}

The sequence of plateaux discussed in previous sections is a consequence of long-lived metastable states in individual realisations. Eventually a finite system leaves each of these states, and proceeds to the next plateau and finally to absorption. In this section, we now seek to engineer a MSVM in which these intermediate states are stable indefinitely. We do this by adding zealots to the population, that is, agents who do not change their opinion state during the dynamics ~\cite{mobilia3, mobilia4, khalilz}. If such agents are added for more than one opinion state, then the system no longer has any absorbing states, and the dynamics continues indefinitely. The question we address here is if and how a configuration of zealots can be chosen so as to stabilise the meta-stable states in Sec.~\ref{sec:individual_realisations}.

\subsection{Competing zealots}

\begin{figure}[h]
\centering
\includegraphics[width=0.95\columnwidth]{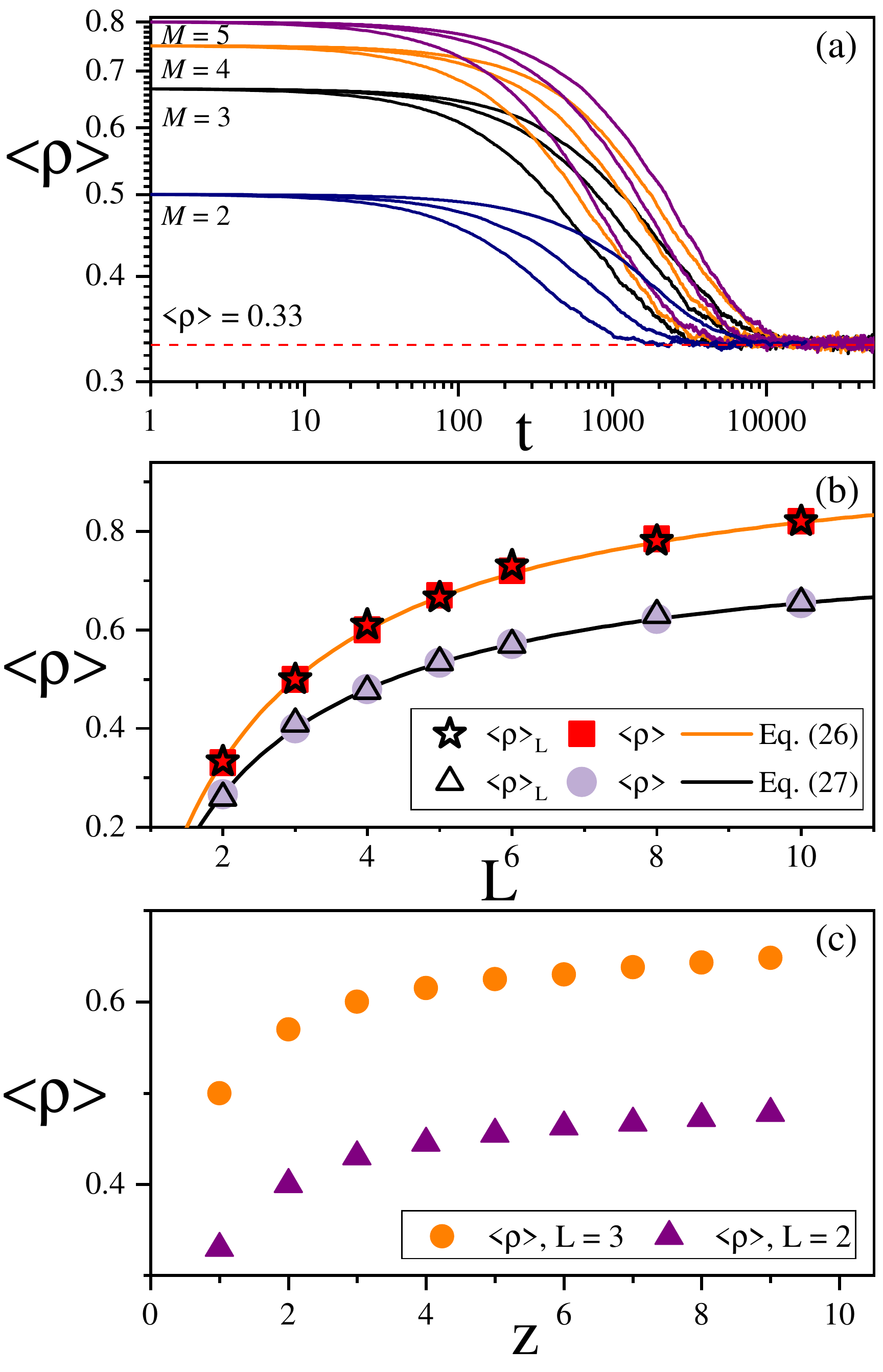}
\caption{(a) Time evolution of the average density of active links (complete graph) when there are two zealots of different opinions. At fixed number of opinions $M$, the size of the population was varied ($N=2000,4000,$ and $6000$). The horizontal dashed line indicates the value $\avg\rho = 0.33$ obtained from setting $L=2$ in Eq.~(\ref{eq:rho_L}).  (b) Filled symbols show the stationary density of active links, $\avg{\rho}$, in a model with one zealot in each of $L$ different opinion states. Squares are for a complete graph, circles for ER networks. Open symbols show the density $\avg{\rho}_L$ in a model with initially $M=15$  states (and no zealots). Stars are for a complete graph, triangles for ER networks. The lines are from Eqs.~(\ref{eq:rho_L}) and (\ref{eq:rho_L_graph}) respectively.  (c) Simulation results for the stationary density of active links in a model with $L$ opinions and $z$ zealots in each of these opinions. Simulations are on complete graphs with system size $N = 4000$.  \label{fig:zealots} }   
\end{figure}

We consider a population of $N$ conventional agents and $Z$ zealots. As before $n_\alpha$ is the number of agents in each opinion state $\alpha = 1, 2, \dots, M$ (not including zealots). We write $z_\alpha$ for the number of zealots for opinion state $\alpha$. Regular agents can change their opinion by interacting with another regular agent or a zealot. Zealots never change their opinion. 

We first focus on the model on a complete graph. The transition rates for this model are then
\be
{\cal T}_{\alpha\to\beta}= \frac{n_\alpha(n_\beta+z_\beta)}{N+Z}
\ee

Suppose now, we start the VM with initially $M$ states, and with two zealots of two different opinions. The remaining $M-2$ opinions will go extinct eventually. In Fig.~\ref{fig:zealots}(a) we show the time evolution for $\avg\rho$ for this system. The stationary density of active links is found as $\avg{\rho}\approx 0.33$ in simulations, irrespective of the initial number of opinions $M$. This density of active links is consistent with that at the plateau for $L=2$ in the model without zealots in the previous section [Eq.~(\ref{eq:rho_L})].

If, more generally, we populate $L$ opinions with one zealot each, then, as seen in Fig.~\ref{fig:zealots} (b), the stationary density of active links also agrees with the plateau value $\avg{\rho}_L$ in the model without zealots. This is found both on the complete graph and on uncorrelated networks.

The data in panel (c) of Fig.~\ref{fig:zealots} shows that the agreement in (a) and (b) between the density of active links of the model with one zealot in each of $L$ opinions and the plateau $\avg{\rho}_L$ in a zealot-free model with initially $M>L$ opinions only holds when there is precisely one zealot in each of the $L$ opinions. We further corroborate this in the next two subsections.

\subsection{Flat stationary distribution of opinions when there is one zealot in each opinion state} \label{app:b}

\subsubsection{Analytical treatment of the model on a complete graph}
We  can demonstrate analytically that the stationary distribution [in the space $(n_1,\dots, n_L)$] of the model with $z$ zealots in each of $L$ opinion states on the complete graph is flat if and only if $z=1$. To do this we consider the master equation
\BE\label{eq:masterZ}
\frac{dP(\textbf n )}{dt}&=&\sum_{\alpha \neq  \beta} P(n_\alpha+1,n_\beta-1) \frac{(n_\alpha+1)[(n_\beta-1)+z]}{N+Z} \nonumber \\
 && - \sum_{\alpha \neq  \beta} P(\textbf n )\frac{n_\alpha(n_\beta+z)}{N + Z},
\EE
where the notation $P(n_\alpha+1,n_\beta-1)$ is a shorthand for $P(E_\alpha E_\beta^{-1}\bn)$, with the raising operator $E_\alpha$ defined in Sec.~\ref{sec:simplify_CG}. Direct algebra shows that $P(\bn)=\mbox{const.}$ is a stationary solution of Eq.~(\ref{eq:masterZ}) if and only if $z=1$.  

\subsubsection{Numerical evidence for the model on networks}
\begin{figure}
\centering
 
\includegraphics[width=1\columnwidth]{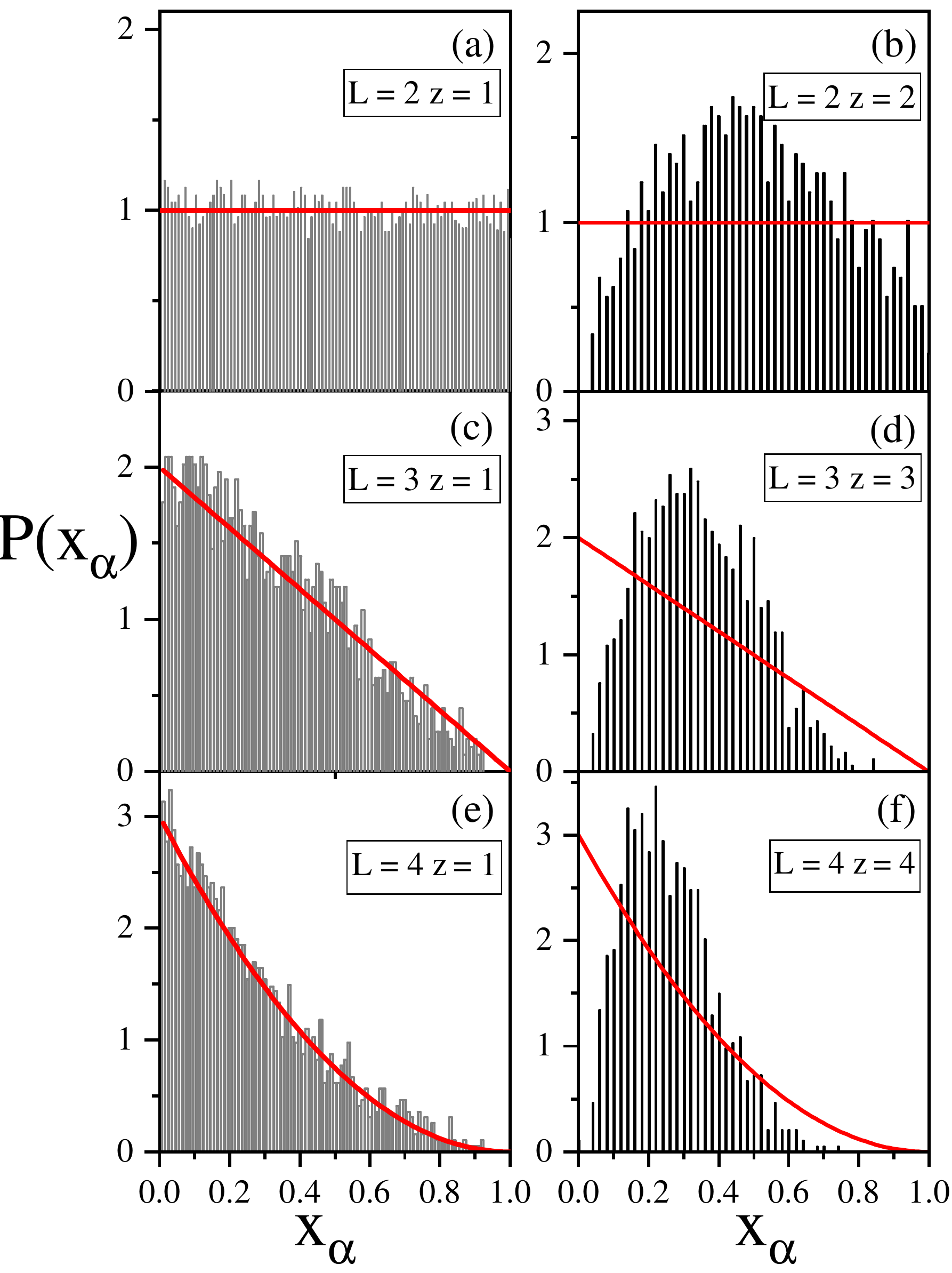}
\caption{Marginal distribution $P_L(x_\alpha)$ for the fraction of agents in any one opinion for a MSVM with $z$ zealots in each of $L$ opinions.  Gray bars are from numerical simulations for ER graphs of size $N=1000$ averaged over $5000$ realisations. The solid lines are the prediction of Eq.~(\ref{eq:plx_main}), derived from the assumption of a flat stationary distribution.}\label{fig:marginal_zealots_net} 
\end{figure}

While an analytical solution for the model on networks is not easily available, simulation results are consistent with the assertion that placing one zealot in each opinion state leads to a flat stationary distribution in the space of $(n_1\dots,n_L)$. We show the marginal stationary distribution $P(x_\alpha)$ for the model with $z$ zealots in each of $L$ opinions in Fig.~\ref{fig:marginal_zealots_net}. For $z=1$ these marginals are well described by the expression in Eq.~(\ref{eq:plx_main}), which is in turn derived from the assumption of a flat distribution $P(x_1,\dots,x_L)$. Panels (b), (d), and (f)  on the other hand demonstrate that the marginals are markedly different when $z\neq 1$.

\section{Summary and conclusions} \label{sec:conclusions}
 In summary, we have investigated the approach of multi-state voter models to consensus. We have distinguished between the behaviour of individual realisations and that of the ensemble average, and we have compared descriptions in terms of local and global properties. We have described local order by the density of active links, and global order by a time-dependent entropy. On networks, we find that local ordering can occur without global ordering. There is an initial coarsening process up to a time $t^\star$, in which the average density of active links decays to a value $\xi(M,\overline k)$ on networks, while the average entropy remains constant. The time scale $t^\star$ is independent of the size of the system. This initial decay is not observed in a complete graph where there is no distinction between local and global order.

At the level of an ensemble average, we further find that the density of active links decays exponentially after time $t^*$, with a time constant $\tau$ which diverges with system size and which is independent of the number of initial opinions. This is observed both on complete graphs and on uncorrelated networks. As in the model with two opinion states, the amplitude and time scale of the exponential decay depend on the structure of the graph. The amplitude and the time scale can be characterised analytically using a pair approximation combined with a reduction to an effective two-state dynamics. 
 
The behaviour of the average entropy is more intricate. We find non-exponential behaviour up to approximately the time at which only two opinions survive. For larger times the entropy decays exponentially with the same time constant $\tau$ as the density of active links. All time scales associated with the decay of the average entropy  diverge with the system size. The time scale $t^\star$ (which is of order $N^0$) has no particular significance for the average entropy. For large systems, the average entropy therefore remains at its initial value for very long times. On networks this is in contrast with the density of active links, which remains at a value associated with partial local ordering and which is lower than the initial density of active links.

 Individual realisations undergo a sequence of extinctions of opinion states, akin to a `jump process'. We find that each realisation remains in plateau values of the active links and entropy between successive extinctions. The location of these plateaux varies from realisation to realisation, and is determined by the configuration of the system at the time of the preceding extinction. The non-equilibrium ensemble average does not provide a proper description of this process, the individual realisations are not self-averaging.  
 
 To describe the ordering process we therefore introduce restricted ensembles of realisations which, at a fixed time, have a given number of surviving opinions. Averaging over these restricted ensembles in simulations allows us to study the mean plateau values exhibited by individual realisations. The average location of the plateaux can also be estimated analytically. The picture we develop is consistent with a maximum spread of configurations at the time of the intermediate extinctions. Using results from Ref.~\cite{starnini} for the mean time between extinctions we are also able to recover the exponential decay of the average density of active links from the jump process for individual realisations.
 
 The restricted ensembles are a useful concept, and allow us to identify partially ordered states along the way to consensus. These states are very long lived in large populations, but eventually the system exits from these states. Arguably, the construction of these ensembles of realisations with $L$ opinions in the system is also somewhat artificial. We have however shown that these partially ordered states can be engineered as genuine stationary states. This is achieved through the introduction of precisely one zealot in each of the $L$ opinions.  

 In closing, we think that the combination of global and local measures of order provides an interesting way of looking at the coarsening dynamics in models of opinion formation. Our work also highlights that the time-evolution of individual realisations can be very different from that of the ensemble average. Randomness can significantly influence individual trajectories, and the effects can last for substantial amounts of time. In other words, `history is contingent' \cite{blount}. The average path to consensus by opinion extinction is therefore of limited use for the description of a given historical empirical occurrence.

\section*{Acknowledgements}

We acknowledge funding from the Spanish Ministry of Science, Innovation and Universities, the Agency AEI and FEDER (EU) under the grant PACSS (RTI2018-093732-B-C22), and the Maria de Maeztu program for Units of Excellence in R\&D (MDM-2017-0711). We acknowledge enlightening discussions with Ra\'ul Toral.


\begin{appendix}   
\section{Normalisation and marginals of the distribution $P_L(x_1,\dots, x_L)$} \label{app:simplex} 
\subsection{Simplex in $K$ dimensions}
For any $K\in\mathbb{N}$ and $y\geq 0$, we define
\be
{\cal S}_{K}(y)=\{(x_1,\dots,x_{K}):x_\alpha\geq 0, \sum_{\alpha=1}^{K} x_\alpha\leq y\},
\ee
as the simplex of size $y$ in $K$ dimensions. We define its volume in $K$-dimensions
\be
V_K(y)=\int_{{\cal S}_{K}(y)} dx_1\dots dx_{K},
\ee
and we have $V_K(y)=y^K V_K(1)$ by simple scaling.

To evaluate $V_K(1)$ we note
\BE
V_K(1)&=&\int_0^1 dx_1 \int_{S_{K-1}(1-x_1)} dx_2\dots dx_L \nonumber \\
&=& V_{K-1}(1)\int_0^1 dx_1 (1-x_1)^{K-1} \nonumber \\
&=&\frac{V_{K-1}(1)}{K}.
\EE
We also have $V_1(1)=1$. By induction therefore, $V_K(1)=1/K!$, and
\be\label{eq:vky}
V_K(y)=\frac{y^K}{K!}.
\ee
\subsection{Normalisation of $P_L(x_1,\dots,x_L)$}
We start from the definition
\be
P_L(x_1,...,x_L) = A\, \delta(x_1 + ... + x_L-1),
\ee
where the $x_\alpha$ are required to be non-negative, and where $A$ is the appropriate normalisation constant. Using $\int dx_1\dots dx_L P_L(x_1,\dots,x_L)=1$, we have after integrating over $x_L$,
\BE
A^{-1}&=&\int_{{\cal S}_{L-1}(1)} dx_1\dots dx_{L-1}\nonumber \\
&=& V_{L-1}(1) = \frac{1}{(L-1)!}.\label{eq:norm}
\EE
Therefore, $A=(L-1)!$.

\subsection{Single-variable marginal of $P_L(x_1,\dots,x_L)$}
We now calculate the single-variable marginal of $P_L(x_1,\dots,x_L)$,
\be
P_L(x_1)=\int dx_2\dots dx_{L} P_L(x_1,\dots,x_L).
\ee
Using the normalisation constant in Eq.~(\ref{eq:norm}) we have
\BE
P_L(x_1)&=&(L-1)! \times  \nonumber \\
&&\int dx_2\dots dx_{L}\, \delta(x_1+x_2+\dots+x_L-1) \nonumber \\
&=&(L-1)! \int_{{\cal S}_{L-2}(1-x_1)} dx_2\dots dx_{L-1} \nonumber \\
&=& (L-1)! V_{L-2}(1-x_1).
\EE
 Hence, using Eq.~(\ref{eq:vky}), we find
\BE\label{eq:plx}
P_L(x_1)&=&(L-1)!\times V_{L-2}(1-x_1) \nonumber \\
&=& (L-1) (1-x_1)^{L-2}.
\EE

\section{Extinction and consensus times in all-to-all geometries and on graphs}\label{app:ext}
In Sec.~\ref{sec:connect} of the main text we use a result for the average consensus time of the multi-state voter model with $K$ opinion states on an all-to-all geometry, and when initial conditions are chosen at random with flat distribution from the simplex defined by $\sum_{\alpha=1}^L x_\alpha=1$. Specifically,
\be\label{eq:TNK}
\avg{T_N(L)}=N\frac{L-1}{L}.
\ee
This was previously reported in Starnini {\em et al.} \cite{starnini}. Starnini {\em et al.} derive this from the backward Fokker-Planck equation of the multi-state model (valid in the limit of large, but finite $N$), and using a separation ansatz exploiting the exchange symmetry between the different opinion states.

We use this consensus time to obtain the mean time, $\avg{t}_{L\to L-1}$ that elapses in a model with initially $L$ opinions until the first extinction, assuming again a uniform distribution of initial conditions in the simplex $\sum_{\alpha=1}^L x_\alpha=1$.
 
 The main purpose of this appendix is to justify the generalisation of the expression for the consensus time $\avg{T_N(L)}$ to uncorrelated networks, Eq.~(\ref{eq:TNK_network}). For convenience we repeat this expression here, 
\be\label{eq:TNK_networks_app}
\avg{T_N(L)}=N\frac{L-1}{L} \times\frac{(\overline k-1)\overline{k}^2}{(\overline{k}-2)\overline{k^2}}.
\ee

Our argument is based on two principles: (a) $\avg{T_N(L)}$  can be obtained from the so-called `mean conditional consensus time' of a suitable two-state model. This applies both in the case of all-to-all interactions and on networks. (b) The mean conditional consensus time for the two-state model in turn can be calculated (in the limit of large, but finite $N$) from the backward Fokker--Planck equation (BFPE) of the respective model. It is known that the Fokker--Planck equation for the two-state model (and hence also the BFPE) on an uncorrelated network can be obtained from that for the model with all-to-all interactions by a re-scaling of time by a factor of $\frac{(\overline k-1)\overline{k}^2}{(\overline{k}-2)\overline{k^2}}$ \cite{vazquez_eguiluz}. This is also discussed in \cite{sood_redner, sood} although the factor is slightly different as explicitly acknowledged in \cite{vazquez_eguiluz}.
 
 We now address (a) and (b) in turn.
 
 \subsection{Reduction to two-state model}
Suppose the dynamics of the $L$-state model is started from an initial condition $\bx=(x_1,\dots,x_L)$ (with $\sum_\alpha x_\alpha=1$) at time $t=0$. Writing $p_C(\bx,t)$ for the probability that consensus (on any opinion) has been reached by time $t$, we have
\be\label{eq:aux2}
p_C(\bx,t)=\sum_{\alpha=1}^L f_\alpha(\bx,t),
\ee
where $f_\alpha(\bx,t)$ is the probability that consensus on opinion $\alpha$ occurs by time $t$. We note that the events on the right in Eq.~(\ref{eq:aux2}) are all mutually exclusive. 

The arrival time distribution at consensus (on any opinion) is then $d p_C(\bx,t)/dt$, and the density of arrivals (per time) at consensus on opinion $\alpha$ is $df_\alpha(\bx,t)/dt$. 

We note that 
\be
q_\alpha(\bx)\equiv \int_0^\infty dt' f_\alpha(\bx,t')
\ee
is the probability that consensus occurs in opinion $\alpha$ (as opposed to another opinion). We have $\sum_\alpha q_\alpha=1$ (consensus occurs with certainty eventually), and in general $q_\alpha<1$ for any one opinion $\alpha$ (i.e., the $f_\alpha$ are not normalised probability densities).

We are interested in mean arrival times, so we calculate
\BE
\avg{T_N(L,\bx)}&\equiv&\int_0^\infty dt'~ t'\frac{d}{dt'} p_C(\bx,t') \nonumber \\
&=&\sum_\alpha \int_0^\infty dt' ~t'\frac{d}{dt'} f_\alpha(\bx,t').
\EE
This is the mean time to consensus (on any opinion). We have explicitly indicated the starting point $\bx$. The average in this expression is only over realisations of the dynamics, but not over the starting point. 

We next note that the mean consensus time, conditional on arrival at consensus on $\alpha$, is given by
\be
\avg{T_\alpha(\bx)}=\frac{\int_0^\infty dt' ~t'\frac{d}{dt'} f_\alpha(\bx,t')}{q_\alpha(\bx)}.
\ee
(we suppress the dependence on $L$ and $N$). Hence,
\BE\label{eq:sum}
\avg{T_N(L,\bx)}&=&\sum_\alpha q_\alpha(\bx) \avg{T_\alpha(\bx)}.
\EE
Consensus occurs at $\alpha$ with probability $q_\alpha(\bx)$, and given $\alpha$ the mean time for this consensus is $\avg{T_\alpha(\bx)}$. 

The key observation is now that, similar to the argument in Sec.~\ref{sec:pa_subsec}, $f_\alpha(\bx,t)$ (and hence also $q_\alpha(\bx)$) can be obtained from looking at a two-state version of the model, in which one opinion is $\alpha$ and where all other opinions are amalgamated into one second opinion state. This idea was used in the context of the voter model in Ref.~\cite{redner} and later in Ref.~\cite{herrerias}. Similar principles had previously been proposed for multi-allele models in the field of genetics \cite{kimura,littler}.

We note that the above argument applies for all-to-all interaction and for networks. We did not make any assumptions on the interaction network in deriving any of the relations up to and including Eq.~(\ref{eq:sum}). What does change in going from all-to-all interaction to networks is the functional form of object such as $p_C(\bx,t)$ and $f_\alpha(\bx,t)$, but not the relations between these quantities.
\medskip

{\em All-to-all interaction.} As an illustration we now show how the reduction to a two-state model can be used to calculate the mean conditional consensus times $ \avg{T_\alpha(\bx)}$, and from these, to derive Eq.~(\ref{eq:TNK}) (valid for all-to-all interactions). 

If the two-state model is started with a proportion $x$ of agents in opinion $1$ (and $1-x$ in opinion $2$), then the mean consensus time conditioned on consensus in opinion $1$ is
\be\label{eq:t_cond}
 \avg{T_1(x)}=-N\frac{1-x}{x}\ln(1-x).
 \ee
This is a well-known result, see for example \cite{sood}. 
 
Using this, and the reduction of the multi-state model to the two-state model we then have
\be\label{eq:t_alpha}
 \avg{T_\alpha(\bx)}=-N\frac{1-x_\alpha}{x_\alpha}\ln(1-x_\alpha).
 \ee
 
We also note that $q_\alpha(\bx)=x_\alpha$ (this follows from the fact that the dynamics of the model preserves the time-average $\avg{x_\alpha(t)}$). 
 
 Hence, inserting into Eq.~(\ref{eq:sum}), 
 \be\label{eq:sum2}
 \avg{T_N(L,\bx)}=-N\sum_{\alpha=1}^L (1-x_\alpha)\ln(1-x_\alpha)
 \ee
 
 We next use this to derive Eq.~(\ref{eq:TNK}). To do this we need to average over $\bx$, with a flat measure in the simplex defined by $x_\alpha\geq 0$ and $\sum_\alpha x_\alpha=1$. Exploiting the symmetry of the model we find from Eq.~(\ref{eq:sum2}),
 \be
 \avg{T_N(L)}=-N L \int_0^1 dx~P_L(x) (1-x)\ln(1-x),
 \ee
 with the marginal $P_L(x)=(L-1)(1-x)^{L-2}$ in Eq.~(\ref{eq:plx_main}). Carrying out the integral we have
 \be
 \avg{T_N(L)}=N\frac{L (L-1)}{L^2}=N\frac{L-1}{L},
 \ee
 i.e., we recover Eq.~(\ref{eq:TNK}).
 \subsection{Extension to uncorrelated networks}
 We now discuss the extension of Eq.~(\ref{eq:TNK}) to uncorrelated networks, i.e., we derive Eq.~(\ref{eq:TNK_networks_app}). All results apply in the pair approximation.
 
Based on the reduction argument in the previous section of this Appendix it is sufficient to show that the mean conditional consensus time of the two-state model in Eq.~(\ref{eq:t_alpha}) generalises to
 \be\label{eq:t_alpha_network}
  \avg{T_\alpha(\bx)}=-N\frac{(\overline k-1)\overline{k}^2}{(\overline{k}-2)\overline{k^2}}\frac{1-x_\alpha}{x_\alpha}\ln(1-x_\alpha).
  \ee

  This time, $\avg{T_\alpha(\bx)}$, in turn is derived from the backward Fokker--Planck equation describing the model in the limit of large, but finite $N$. This is standard in the case of all-to-all interaction (see e.g. \cite{sood}).
  
 The crucial observation is now that the (forward) Fokker-Planck equation of the two-state model on uncorrelated graphs is obtained from that for the two-state model with all-to-all interactions by a simple re-scaling of time by a factor $[(\overline k-1)\overline{k}^2]/[(\overline{k}-2)\overline{k^2}]$. This is discussed in \cite{vazquez_eguiluz}, see for example Eq.~(12) in this reference (see also \cite{sood, sood_redner}).
 
 The same re-scaling then also applies to the backward equation, and it then follows immediately that all time scales derived from the backward equation also undergo the same re-scaling. This then leads to Eq.~(\ref{eq:t_alpha_network}).

\end{appendix}

\end{document}